\documentclass[epjST_arxiv]{svjour_arxiv}
\usepackage{graphicx} 
\usepackage{epstopdf} 
\usepackage{epsfig} 
\usepackage{subfigure} 
\usepackage{amsmath,amsfonts,amssymb}
\usepackage{mathbbol}
\usepackage{braket}
\usepackage{ucs}		
\usepackage[utf8x]{inputenc}	
\usepackage[T1]{fontenc}	
\usepackage[english]{babel}	
\usepackage{verbatim}		
\usepackage{float}
\usepackage{color}
\usepackage{indentfirst}		

\usepackage[squaren]{SIunits}
\DeclareRobustCommand{\@NewUnit}[2]{\addunit{#1}{#2}} 
\@NewUnit\gauss{G} 
\newcommand{\eff}{ef\hspace{-0.5mm}f}
\newcommand{\OD}{O\hspace{-0.5mm}D}
\definecolor{orange}{RGB}{230,105,0}
\usepackage[square,sort,comma,numbers,compress]{natbib} 

\usepackage{hyperref}
\hypersetup{
    colorlinks=true,
    linkcolor=blue,
    filecolor=blue,      
    urlcolor=blue,
    citecolor=blue,
    pdftitle={IEI of Rydberg P states},
}

\begin{document}
\title{Interaction-Enhanced Imaging of Rydberg P states}
\subtitle{Preparation and detection of Rydberg atoms for engineering long-range interactions}
\author{Vladislav Gavryusev\inst{1}\fnmsep\thanks{\email{gavryusev@physi.uni-heidelberg.de}} \and Miguel Ferreira-Cao\inst{1} \and Armin Keki\'{c}\inst{1} \and Gerhard Z{\"u}rn\inst{1} \and Adrien Signoles\inst{1}}
\institute{Physikalisches Institut, Universit{\"a}t Heidelberg, Im Neuenheimer Feld 226, 69120, Heidelberg}
\abstract{
The Interaction Enhanced Imaging technique allows to detect the spatial distribution of strongly interacting impurities embedded within a gas of background atoms used as a contrast medium~\cite{Guenter2012}. Here we present a detailed study of this technique, applied to detect Rydberg $P$ states. We experimentally realize fast and efficient three-photon excitation of $P$ states, optimized according to the results of a theoretical effective two-level model. Few Rydberg $P$-state atoms, prepared in a small cloud with dimensions comparable to the blockade radius, are detected with a good sensitivity by averaging over 50 shots. The main aspects of the technique are described with a hard-sphere model, finding good agreement with experimental data. This work paves the way to a non-destructive optical detection of single Rydberg atoms with high spatial and temporal resolution.
} 
\maketitle
\setcounter{tocdepth}{2}
\tableofcontents
\section{Introduction}
\label{sec:intro}

Ultracold Rydberg atoms with their strong and long-range interactions \cite{Saffman2010} offer numerous possibilities to study equilibrium and non-equilibrium properties of strongly correlated matter \cite{Hazzard2014}. Over the last few years we have witnessed an explosion of interest in these systems for studying diverse phenomena including the formation of spatially correlated crystalline-like structures~\cite{pohl2010,Schauss2012,Levi2015} or ultracold plasmas~\cite{Vanhaecke2005,Vrinceanu2009,Pohl2011}, the emergence of extreme nonlinear optical effects~\cite{Pritchard2010,Peyronel2012,Firstenberg2013} and the transport of energy through Rydberg aggregates~\cite{Schonleber2015}. However, until recently, optical techniques for state-resolved detection of Rydberg atoms with high spatial and temporal resolution have been missing. In this paper we present a detailed study of a technique called Interaction Enhanced Imaging (IEI)~\cite{Guenter2012}, that we apply to the state resolved detection of Rydberg $P$-states (with angular momentum $L=1$) embedded in a gas of atoms coupled to Rydberg $S$-states ($L=0$). This state combination is particularly relevant for time resolved studies of coherent and incoherent transport of Rydberg excitations mediated by the strong dipolar exchange interactions~\cite{Schonleber2015,Westermann2006,Schempp2015}.

In this introduction, we give an overview of the detection method and compare it to state-of-the-art cold Rydberg gas experiments. In section \ref{sec:ModelIEI}, we present an intuitive model that captures the main physical principles of the imaging technique and provides an estimation of the achievable signal-to-noise ratio. Section \ref{sec:setup} describes our experimental setup, while in section \ref{sec:Excitation} we show how we prepare Rydberg atoms in a P-state. Finally, the first results on the optical detection of Rydberg P-states using IEI are presented in section \ref{sec:IEI_experimental}.\\

\noindent \emph{State-of-the-art of Rydberg atom imaging}\\

So far there are only a few techniques suited to spatially image Rydberg atoms. An early example is based on field ion microscopy demonstrated by Schwarzkopf \emph{et al.}~\cite{Schwarzkopf2011}. Here Rydberg atoms excited from a magneto-optical trap are ionized by applying a large voltage onto a needle with a rounded tiny tip, which creates a strong inhomogeneous electric field. Following the divergent field lines, the resulting ions are guided towards a fluorescent multichannel plate (MCP). Each ion is revealed by a bright spot of light onto a charge-coupled-device (CCD) camera, allowing for a resolution of the order of a few $\micro\meter$. Nevertheless, the detection efficiency is limited to $\leq 50\%$ by the quantum efficiency of the MCP. Using this technique, spatial correlations through the pair correlation function and van der Waals interactions between Rydberg atoms have been measured~\cite{Schwarzkopf2013,Thaicharoen2015,Weber2014,Manthey2015}. 

Spatially resolved detection of Rydberg atoms trapped in an optical lattice has also been achieved using an optical approach~\cite{Schauss2012}. Rydberg atoms are de-excited via stimulated emission by laser light towards a short-lifetime intermediate state and are captured in an optical lattice. Their position is then revealed by high-resolution fluorescence imaging using a lower transition. With this technique, strong spatial correlations have been observed with nearly single-site resolution and a detection efficiency of $\sim 75\%$. A similar technique allows to detect Rydberg atoms trapped in small arrays with a nearly $97\%$ efficiency~\cite{Labuhn2016}. Although the detection sensitivity is very good, this technique is best suited to two-dimensional lattices or optical tweezers and the detection method destroys the Rydberg atoms.

Other methods have been introduced for alkaline-earth atomic species, taking advantage of the state-dependent optical transition of the core electron to probe the presence of Rydberg atoms. By shining laser light resonant to a transition of the core electron in Strontium with one electron excited in a Rydberg state, the fluorescence light can be measured, revealing the presence of Rydberg atoms~\cite{McQuillen2013}. The technique provides very good temporal resolution of $\sim 10\,\nano\second$, but low spatial resolution ($\sim 200\,\micro\meter$) and is most sensitive to high-$\ell$ states. For low-$\ell$ states, a more sophisticated technique allowing for larger spatial resolution consists in exciting the core electron to an intermediate state, which leads with a large probability to autoionization of the Rydberg state. Using a focused laser beam, one can then locally ionize the atoms and record the Rydberg distribution with a spatial resolution of $10\,\micro\meter$~\cite{Lochead2013}. 
\\

\noindent \emph{Review of Interaction Enhanced Imaging (IEI)}\\

\begin{figure}[ht]
\centering
\includegraphics[width=0.8\linewidth,keepaspectratio=true]{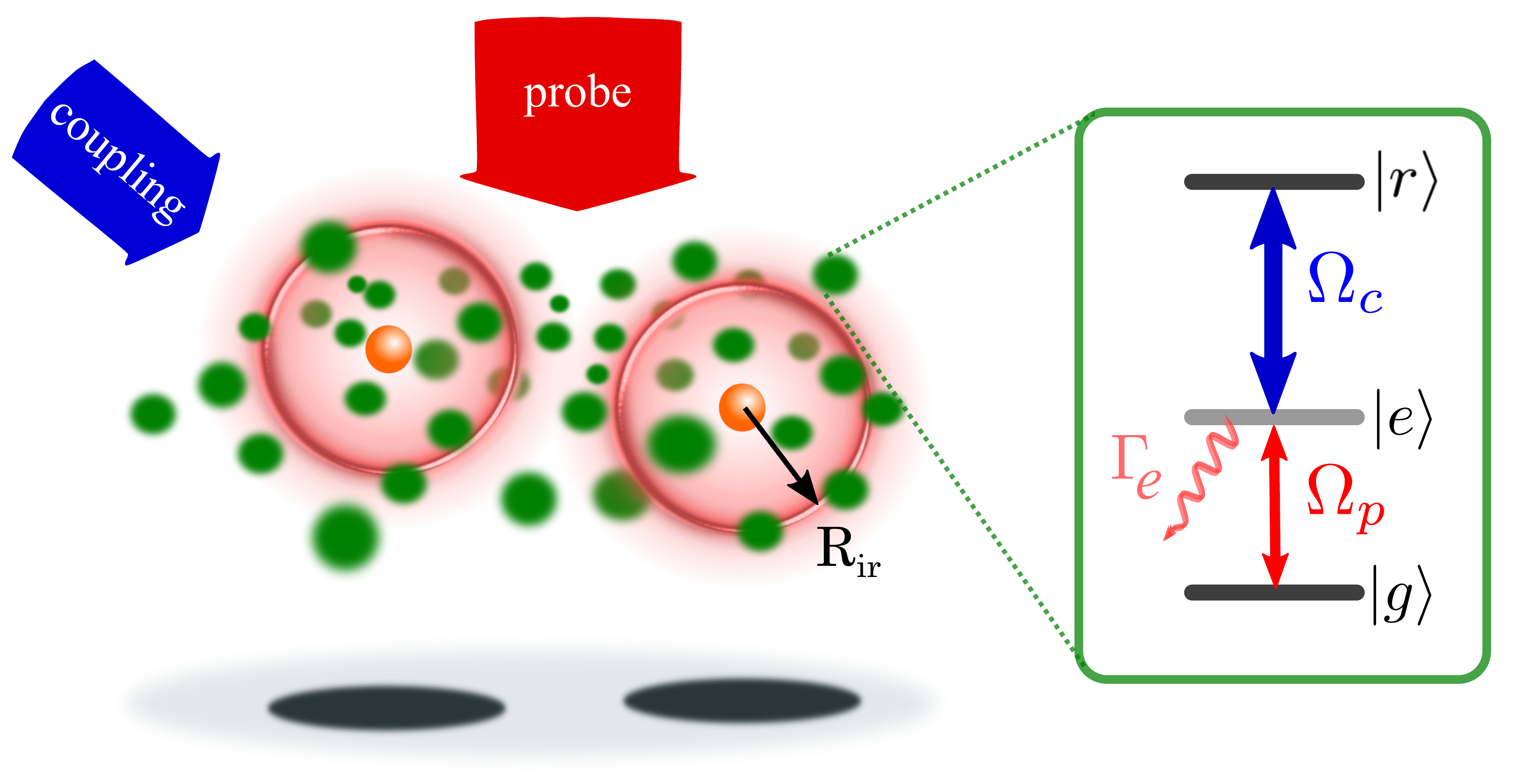}
\caption{Scheme for interaction-enhanced absorption imaging of individual impurities (orange spheres) within a dense gas of probe atoms (green spheres). Two coherent resonant light fields called \emph{probe} and \emph{coupling}, with Rabi frequencies $\Omega_p$ and $\Omega_c$, couple the ground state $\ket{g}$ of the probe atoms to a Rydberg state $\ket{r}$, inducing the transparency condition for the probe light field for atoms far from any impurity. However the impurities and the surrounding atoms interact strongly within a critical distance $R_{ir}$, leading to an energy shift of the Rydberg state $\ket{r}$ which locally breaks the EIT condition. Consequently the probe atoms become absorptive in these regions. The presence of an impurity is revealed by a shadow on the camera which collects the probe light after its propagation through the atomic cloud.}
\label{fig:IEI_sph_scheme}       
\end{figure}

To study dipole-mediated transport dynamics, our group recently implemented a new type of detection method which combines single particle sensitivity and high spatial and temporal resolution. We experimentally implemented an absorption technique to detect Rydberg atoms, so-called impurities, embedded in an atomic gas, with a spatial resolution below the Rydberg blockade radius. This IEI technique, closely based on our original proposal~\cite{Guenter2012} and also related to a similar proposal by Olmos \emph{et al.}~\cite{Olmos2011}, consists in measuring the absorption of a probe light propagating through the gas, that acts as a contrast medium, in such a way that the absorption is modified by the presence of the Rydberg impurities. The method has been demonstrated in our experiment with rubidium Rydberg atoms, but it can be potentially applied for any particle that exhibits strong coupling to a medium, like ions via Coulomb interactions and polar molecules via dipole-dipole interactions. The key to IEI is to use the background atoms to reveal the presence of impurities. To do so, the gas is coupled under electromagnetically-induced transparency (EIT) conditions to an auxiliary "probe" Rydberg state by a weak probe field and a strong coupling field, being therefore transparent to the probe light~\cite{Fleischhauer2005,Pritchard2011,Sevinccli2011}. However, atoms close to an impurity experience dipole-dipole interactions that energetically shift this probe Rydberg state and break the EIT~\cite{Hofmann2013,GarttnerPRA2014}. This effect thus casts a shadow in the spatial absorption profile that allows to map the position of the impurities (fig.~\ref{fig:IEI_sph_scheme}). 


This method exhibits numerous advantages that are of fundamental importance to investigate dynamics of Rydberg excitations in ultracold gases. Since the detection signal is generated from background atoms surrounding the Rydberg impurities, the signal can be strongly enhanced, allowing to detect the impurity with high efficiency. Those background atoms act like an amplifier with an enhancement factor equal to the number of additional absorbers per impurity. In the blockade picture, one can define a critical distance $R_{ir}$ from an impurity below which probe atoms are considered as absorbers. The amplification factor, which can be now defined as the number of probe atoms within the blockade sphere, can be tuned to strongly enhance absorption around a single impurity. Therefore, the detection is potentially single-impurity sensitive, as soon as the detected signal is larger than the imaging noise. One experimental snapshot might be enough to reveal the presence of the impurities with a good spatial resolution. Finally, the detection keeps the number of impurities constant and should allow for multiple measurements similarly to quantum non-destructive measurements~\cite{Grangier1998,Guerlin2007}.
\\


\noindent \emph{Connection to similar experiments}\\

The IEI technique uses the optical absorption spectrum to map out atomic properties. In a complementary point of view, it is possible to use Rydberg gases to modify an optical field. Many studies have been recently done to investigate how dipole-dipole interactions affect propagation of light, for instance leading to non-linear absorption~\cite{Ates2011,Sevinccli2011}, dephasing of Rydberg polaritons~\cite{Tresp2015} and photon-photon interactions~\cite{Gorshkov2011,Peyronel2012}. IEI differs from these experiments because it exploits strong inter-state interactions between the Rydberg impurities and Rydberg polaritons to imprint information on the spatial distribution directly onto the light field.

In Rydberg gases coupled under EIT, the strong change of optical susceptibility due to the presence of impurities opens interesting perspectives to engineer control on the quantum state of optical photons. Recent developments towards a single-photon optical transistor have been done, where a single "control" photon is used to substantially change an optical signal~\cite{Baur2014,Tiarks2014,Gorniaczyk2014}. In practice, a resonant light signal propagates through a gas of Rubidium atoms under EIT condition, where previously a single photon could be stored in the form of a Rydberg polariton~\cite{Fleischhauer2000,Hofmann2013,Maxwell2013}. As in IEI, the transparency of the gas depends on the state ($\ket{0}$ or $\ket{1}$) of the control photon. In this way, the storage of a single control photon affects the propagation of many other photons, constituting a first step in the realization of a photon transistor. Although the purpose is different, these experiments exhibit strong similarities with IEI: the absorption of a signal (resp. probe) light is strongly enhanced by the presence of a Rydberg polariton (resp. impurity) in an EIT medium. The efficiency of the transistor will then be measured by a gain, characterizing the number of signal photons affected by the presence of a polariton. In IEI we define an amplification factor which represents the number of atoms affected by the presence of an impurity. Both can be enhanced by increasing the interaction strength, for instance by tuning two S states into a F{\"{o}}rster resonance \cite{Tiarks2014,Gorniaczyk2015}.\\

\noindent \emph{First experimental realization of IEI}\\

\begin{figure}[ht]
\centering
\includegraphics[width= \linewidth,keepaspectratio=true]{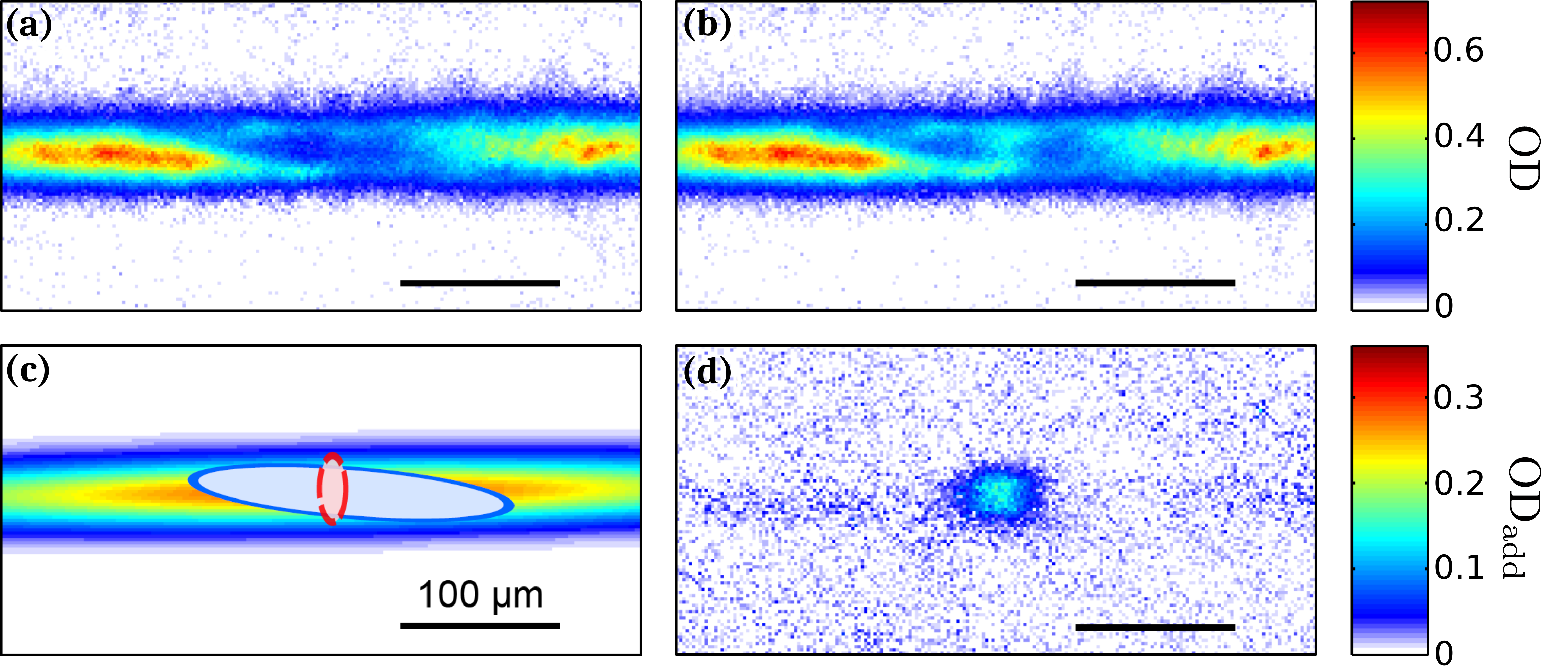}
\caption{Absorption images acquired during an interaction-enhanced imaging experiment (reproduced from~\cite{Gunter2014phd}). {\bf(a)} In a large region in the center of the cloud, atoms are illuminated by a strong coupling beam with $\Omega_c \sim 2\pi \cdot 9\,\mega\hertz$, which renders them transparent to the probe light. {\bf(b)} A second absorption image is acquired with Rydberg impurities pre-excited in the center of the cloud. They locally break the EIT condition, rendering the atoms absorptive again. {\bf(c)} Sketch of the atomic cloud and of the positions of the EIT-coupling (blue) and excitation (red) beams. The ellipses indicate the regions corresponding to $1.5\cdot\sigma$ of the intensity profiles. {\bf(d)} The difference in absorption between (a) and (b) reveals the additional absorbers and contains information about the number and the spatial distribution of the Rydberg impurities. The images are averaged over $150$ repetitions, with an exposure time of $5\,\micro\second$. For this data, we estimated a number of $\sim 300$ impurities.}
\label{fig:IEI_experiment}       
\end{figure}

Interaction Enhanced Imaging has been successfully implemented to measure the spatial distribution of Rydberg impurities in a gas of Rubidium~\cite{Guenter2013}. The atoms are initially prepared in their electronic ground state in a dipole trap at a temperature of few $\micro\kelvin$. They are then illuminated by a large $780\,\nano\meter$ probe laser beam to ensure an uniform probe intensity and with a $480\,\nano\meter$ blue laser beam focused in a smaller region, considered as the region of interest~(fig.\ref{fig:IEI_experiment}c). Together, these light fields coupled the ground state $\ket{5S_{1/2}}$ to the $\ket{37S}$ Rydberg state, inducing an EIT condition on the $\ket{5S} \leftrightarrow \ket{5P}$ transition. Therefore, atoms in this region were rendered transparent to the probe light (fig.\ref{fig:IEI_experiment}a). In a second experimental repetition, impurities in the state $\ket{50S}$ were first excited in the very center of the trap by means of a two-photon transition, then the EIT coupling was turned on. The transparency contrast in this region drops because of the presence of these impurities (fig.\ref{fig:IEI_experiment}b) which break the EIT condition. The final signal is obtained by calculating the difference between the two transmission images, extracting the additional absorption induced by the impurities~(fig.\ref{fig:IEI_experiment}d).

This experiment illustrates how IEI allows to map the distribution of Rydberg impurities embedded in an atomic gas. The technique has been already used to investigate classical diffusion induced by dipole-dipole interactions~\cite{Guenter2013}, making it possible to tackle many questions concerning e.g. the transport mechanisms, the effect of disorder and the role of dissipation. In the experiment the electric field was tuned close to a F{\"{o}rster resonance to induce state changing interactions. To minimize the number of excitation pathways which can influence the transport, we aim to study transport induced by resonant dipolar interaction between $P$ and $S$ atoms which are much stronger than indirect van der Waals exchange. 
Here, we report the first experimental realization of IEI using Rydberg $nP$ states and we present a detailed study of the detection method that allows for enhancement of the sensitivity and of the signal-to-noise ratio (SNR). An analysis of the different processes involved in the imaging, based on a hard-sphere model, is developed in the following section.


\section{Model for Interaction Enhanced Imaging of Rydberg atoms}
\label{sec:ModelIEI}

Interaction-enhanced imaging is based on the different optical responses of an atomic cloud with and without the impurities it is aimed to detect. To determine optimal conditions for imaging we developed a simple model which captures the main features of IEI, especially the impurity-probe interactions and the probe-probe interactions that naturally arise in the gas. This model describes the medium as composed of three inter-dispersed sub-volumes: a first region where transparency is preserved, a second region in which transparency is reduced due to probe-probe interactions and a third one in which transparency is affected by impurity-probe interactions. To account for the relative size of each volume we employ a simple hard-sphere approach to quantify each region.

The atom-light interaction is treated semi-classically and we will consider the system to be in the steady state, since the imaging pulses of few $\rm{\mu s}$ are much longer than the typical decay time $\Gamma_e^{-1}$. Additionally we will assume in the following that the atoms are cooled to temperatures low enough to neglect thermal motion and mechanical effects on the Rydberg atoms~\cite{Amthor2007,Teixeira2015}.

\subsection{Light propagation under ideal EIT conditions}
\label{subsec:EIT0}

Electromagnetically induced transparency is a quantum effect that renders an otherwise absorptive medium transparent. It is especially manifest in atomic systems when an intermediate, fast-decaying level is coupled to two metastable states~\cite{Fleischhauer2005}. In our system, the probe atoms fulfil this condition (see fig.~\ref{fig:IEI_sph_scheme}) since the decay of the Rydberg state $\ket{r}$ is negligible compared to the one of the intermediate state $\ket{e}$. Here we describe the optical response of an atomic cloud under EIT conditions, meaning a weak probe field of Rabi frequency $\Omega_p$ resonant on the $\ket{g} \leftrightarrow \ket{e}$ transition and a strong coupling field of Rabi frequency $\Omega_c$ resonant on the $\ket{e} \leftrightarrow \ket{r}$ transition. Moreover, we assume at this stage that interactions between atoms can be fully neglected, which is reasonably valid in a dilute gas.

Exposing the sample to a probe light field of intensity $I \propto \Omega_p^2$, the optical properties of a gas of density $n$ can be well described through its first order susceptibility $\chi$ defined as
\begin{equation}
	\label{eq:def_chi}
	\chi = \frac{\sigma_0 n \Gamma_e}{k \Omega_p} \rho_{eg}
\end{equation}
where $\sigma_0$ is the resonant absorption cross-section, $k$ the wavevector and $\rho_{eg}$ the single-atom density matrix element for the chosen probe transition $\ket{g} \leftrightarrow \ket{e}$. In our case, the density is a function of position $n(\bf{r})$ due to the Gaussian profile of the atomic cloud, while $\rho_{eg}$ depends in particular on the probe intensity $I$. Under EIT conditions, $\rho_{eg}$ can be expressed analytically, allowing for the calculation of the ideal EIT susceptibility $\chi^0_{eit}$~\cite{Fleischhauer2005}. On resonance, $\chi^0_{eit}$ depends on the effective dephasing rate $\gamma_{gr}/2$ of the coherence density matrix element $\rho_{gr}$, including contributions of the decay rate of $\ket{r}$ and of additional dephasing rates, and is expressed as
\begin{equation}
 \label{eq:chi_EIT0}
 \chi_{eit}^0=\frac{\sigma_0 n}{k} \frac{i\Gamma_e^2}{\Gamma_{e}^2+\Omega_c^2\Gamma_e/\gamma_{gr}+2\Omega_p^2}.
\end{equation}
Here we suppose that the sources of dephasing acting on the intermediate level $\ket{e}$ are negligible compared to $\Gamma_e$, which is experimentally verified (sec. \ref{subsec:SetupExcDet}). Equation~\eqref{eq:chi_EIT0} shows that $\chi_{eit}^0 \ll \sigma_0 n/k$ as soon as $\Omega_c^2 \gg \Gamma_e \gamma_{gr}$, which is satisfied in our experiment for coupling Rabi frequencies $\Omega_c$ on the order of few $\rm{MHz}$. Since the absorption is related to the imaginary part of the susceptibility, such conditions lead to transparency of the cloud for the probing beam.

\subsection{Inclusion of probe-probe interactions}
\label{subsec:EIT}

The validity of equation~\eqref{eq:chi_EIT0} breaks down as soon as the Rydberg atomic density is large enough such that the van der Waals interaction between the Rydberg states of the probe atoms becomes significant. Experimentally, the change in transparency as function of Rydberg density has been already observed in Rydberg media~\cite{Sevinccli2011,Pritchard2010,Pritchard2011,Hofmann2013,DeSalvo2016,Han2016}. Moreover it has been shown from a semi-analytic rate equation (RE) approach that the effect of these interactions can be approximated by an energy shift $\Delta_{int}=V_{rr}$ of the Rydberg state when we want to estimate the optical response of the gas~\cite{GarttnerPRA2014,Ates2011} (fig.~\ref{fig:IEI_HS_scheme}c). This shift effectively introduces a detuning $\Delta_c=\Delta_{int}$ of the coupling light that modifies the effective dephasing $\gamma_{gr} \to \gamma_{gr}+2i\Delta_{int}$ in eq.~\eqref{eq:chi_EIT0}, leading to a loss of transparency. For large energy shift $V_{rr}$ the susceptibility approaches the resonant two-level susceptibility $\chi_{2l}^{~}$ which is given by
\begin{equation}
 \label{eq:chi_2lvl}
 \chi_{2l}^{~} = \frac{\sigma_0 n}{k} \frac{i \Gamma_e^2}{\Gamma_e^2+2\Omega_p^2}.
\end{equation}
The dependence on $\Omega_p$ takes into account the saturation of the atomic transition in contrast to the weak probe limit. In figure~\ref{fig:IEI_HS_scheme}a the ratio between the imaginary part of $\chi_{eit}^0$ normalized by the two-level one is plotted as a function of the single-photon detuning $\Delta_c$ between the coupling field and the atomic transition $\ket{e} \leftrightarrow \ket{r}$. It shows that the transparency is preserved in a window around the resonance with a FWHM $\sigma_{eit} \approx \Omega_c^2/\Gamma_e$ which defines the EIT bandwidth.

Ates \emph{et al.}~\cite{Ates2011} has found using a rate equation model and Monte-Carlo simulations that the optical response including probe-probe interactions can be related to the Rydberg atomic density through the relation
\begin{equation}
 \label{eq:chi_eit}
 \chi_{eit}^{~} = f_{rr} \chi_{2l}^{~} +  \left(1-f_{rr}\right) \chi_{eit}^0,
\end{equation}
where $f_{rr}$ is the effective fraction of volume blockaded by the probe-probe interactions, expressed as
\begin{equation}
 \label{eq:frr}
 f_{rr}=\frac{n\rho_{rr}^0-n\rho_{rr}}{n\rho^0_{rr}} = 1-\frac{\rho_{rr}}{\rho^0_{rr}}.
\end{equation}
Here $\rho_{rr}^0$ is the density matrix element in the non-interacting regime (single atom case, equal to $\Omega_p^2/\Omega_c^2$ in the steady state) and $\rho_{rr}$ the one in the interacting regime. The blockaded volume fraction is represented by the fraction of atoms that are not excited to the Rydberg state $\ket{r}$ because of the probe-probe interactions. Therefore $f_{rr}$ is expressed through the normalized difference between $n \rho_{rr}^0$, the density of Rydberg atoms that would have been excited without interactions, and $n \rho_{rr}^{~}$, the density of Rydberg atoms that are effectively excited. Nevertheless, eqs.~\eqref{eq:chi_eit} and~\eqref{eq:frr} do not allow to calculate the optical susceptibility since they do not provide the expression of $\rho_{rr}$ which requires a self-consistent solution to properly take into account collective effects present in the system.

To determine the fraction $f_{rr}$ we use a hard-sphere approximation~\cite{Petrosyan2011,Sevinccli2011,Parigi2012} in which atoms that experience an energy shift $V_{rr}$ larger than half of the EIT bandwidth $\sigma_{eit}/2$ will be considered as two-level absorbers, while the others are assumed to be coupled under ideal EIT conditions (sec.~\ref{subsec:EIT0}). Hence probe atoms surrounding any given atom experience the interaction only if their distance $r$ is smaller than the probe-probe Rydberg blockade radius $R_{rr}$
\begin{equation}
	\label{eq:Rrr}
	r < R_{rr} = \left( \dfrac{2C_6^{rr}}{\sigma_{eit}} \right)^{\frac{1}{6}},
\end{equation}
where $C_6^{rr}$ is the van der Waals coefficient of the $\ket{r}$ state determining $V_{rr}=C_6^{rr}/r^6$. Within this approach the fraction $f_{rr}$ can be expressed as the number of atoms contained in each blockaded volume $V_{bl}=4/3\cdot \pi R_{rr}^3$ over the total number of atoms, leading to
\begin{equation}
	\label{eq:frr2}
	f_{rr} = n \rho_{rr} V_{bl}.
\end{equation}
By substituting eq.~\eqref{eq:frr2} in eq.~\eqref{eq:frr} we can extract the density matrix element $\rho_{rr}$
\begin{equation}
	\label{eq:rho_rr}
	\frac{\rho_{rr}}{\rho_{rr}^0} = \frac{1}{1+n\rho_{rr}^0 V_{bl}}
\end{equation}
and obtain an analytical expression for the optical susceptibility in the presence of probe-probe interactions.

\begin{figure}[t]
\centering
\includegraphics[width=\linewidth,keepaspectratio=true]{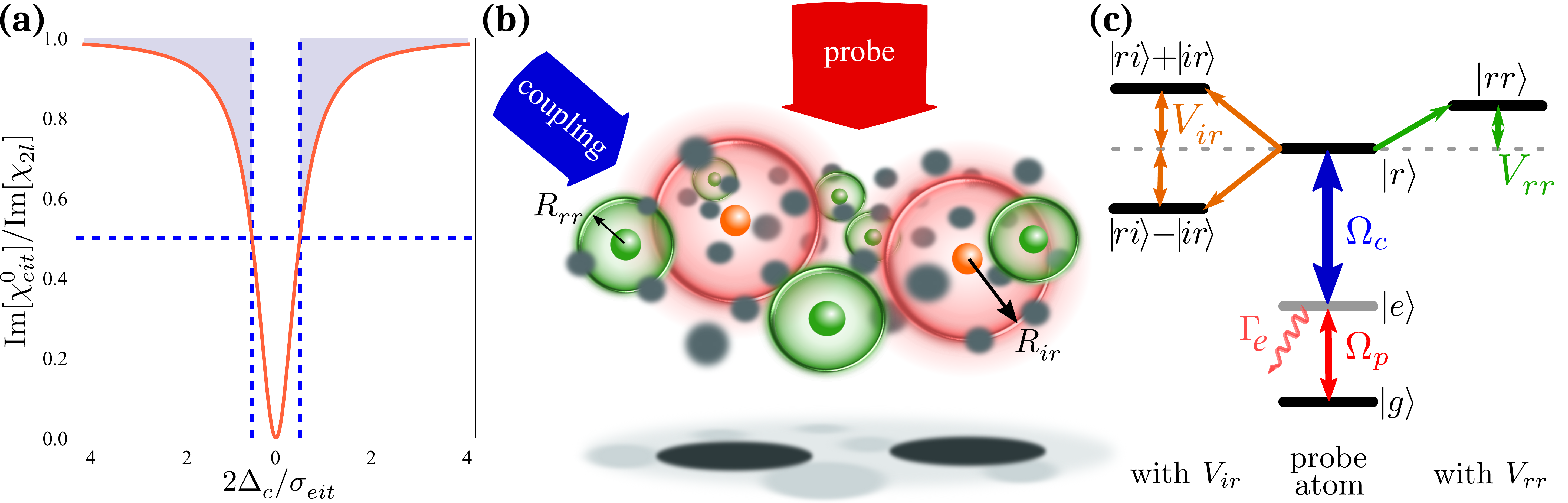}
\caption{Hard-sphere description of IEI. {\bf(a)} Ratio between the imaginary parts of the EIT and two-level susceptibility as a function of the detuning $\Delta_c$ of the $\ket{e} \leftrightarrow \ket{r}$ transition, normalized by the FWHM of the EIT transparency window. At $\Delta_c = \sigma_{eit}/2$ the susceptibility is half of the two-level one. For larger detunings atoms are considered as two-level absorbers (filled area) {\bf (b)} Description of the interactions with a hard-sphere approach: the impurities (orange dots) interact with the surrounding probe atoms (grey and green dots when respectively in $\ket{g}$ or $\ket{r}$). They induce a level shift $V_{ir}$ that, within a distance $R_{ir}$, breaks the EIT condition. The probing laser beam is imaged on a CCD camera and around the position of each impurity an absorption spot is expected. 
Within a distance $R_{rr}$ around each $\ket{r}$ Rydberg state, probe-probe interactions reduce the transparency as well. {\bf(c)}~Without interactions (middle) the three states $\ket{g}$, $\ket{e}$, $\ket{r}$ of the probe atoms are coherently coupled by two resonant light fields with Rabi frequencies $\Omega_p$ and $\Omega_c$. The probe-probe interactions induce a level shift of $V_{rr}$ on $\ket{r}$ (right). Due to the presence of a Rydberg impurity~$\ket{i}$ which interacts through the dipolar exchange with $\ket{r}$, the new eigenstates are split by an energy $2 V_{ir}$ (left). Effectively both types of interaction lead to a vanishing probability to excite $\ket{r}$ when their strength exceeds the EIT bandwidth and to absorption on the $\ket{g} \leftrightarrow \ket{e}$~transition.}
\label{fig:IEI_HS_scheme}       
\end{figure}

\subsection{Effect of impurities}
\label{subsec:modelHS}

We follow a very similar approach to incorporate the optical response in the presence of impurities $\ket{i}$ which have been independently excited. In the specific case of imaging of $P$ Rydberg impurities using $S$ probe atoms, the two states experience a dipole-dipole exchange interaction. The pair eigenstates are the symmetric and antisymmetric combinations $(\ket{ri}\pm\ket{ir})/\sqrt{2}$ of the unbound pair states, split in energy by $\pm \Delta_{int}= \pm V_{ir} = \pm C_3^{ir}/R^3$~\cite{Lukin2001,Comparat2010}, where $C_3^{ir}$ is the interaction strength of the resonant $\ket{i} \leftrightarrow \ket{r}$ dipolar interaction (fig.~\ref{fig:IEI_HS_scheme}b,c). In analogy to the probe-probe interactions we define a critical radius $R_{ir}$
\begin{equation}
 \label{eq:Rir}
 R_{ir} = \left( \dfrac{2 C_3^{ir}}{\sigma_{eit}} \right)^{\frac{1}{3}}
\end{equation}
beyond which the energy splitting is smaller than the EIT bandwidth $\sigma_{eit}$, allowing to address the Rydberg $\ket{r}$ state. Beyond this critical distance we assume the gas to behave as a three level system, while below the susceptibility is approximated by the one of a two level system. Since the interactions are anisotropic and the experiment is performed with 3D random positions of the atoms, we consider an effective isotropic reduced interaction strength $C_{3}^{ir}$ that is obtained by averaging over the total solid angle.

With this hard-sphere model it becomes possible to express the optical susceptibility of a medium under EIT coupling in the presence of impurities as the weighted combination of the two-level susceptibility (eq. \eqref{eq:chi_2lvl}) and the EIT susceptibility with probe-probe interactions (eq. \eqref{eq:chi_eit})
\begin{equation}
 \label{eq:chi_IMP}
 \chi_{imp}^{~} =  f_{ir} \chi_{2l}^{~} + \left(1-f_{ir}\right) \chi_{eit}^{~},
\end{equation}
where $f_{ir}$ is the fraction of volume blockaded by the impurity-probe interactions. At low impurity density $n_i$ this fraction can be estimated by geometric arguments assuming no overlap between the blockade sphere around each impurity. Each impurity blockades a volume $V_{bl,i}=4/3\cdot \pi R_{ir}^3$, so that $N_i$ impurities blockade a total volume $N_i V_{bl,i}$, leading to a blockaded fraction $f_{ir}=n_i V_{bl,i}$. In the opposite case of large impurity densities $f_{ir}$ is equal to 1 since the whole volume is blockaded. In analogy to section~\ref{subsec:EIT} we extrapolate the relation to $f_{ir}= 1-1/(1+n_i V_{bl,i})$.

\subsection{Detection method, signal and noise sources}
\label{sec:detection}

To probe the impurity distribution we measure the absorption of the probe light induced by the atomic cloud. The light is collected by a CCD camera with high-sensitivity that allows for a fast, spatially-resolved measurement. Assuming a paraxial approximation, the propagation of a stationary probe field with wavevector $k$ through the medium along the $x$ direction is described by the differential equation
 \begin{equation}
  \label{eq:diff_light_prop}
  \frac{\partial I(x)}{\partial x} = i k \chi \Big( I(x),n(x) \Big) I(x).
 \end{equation}
In general $\chi$ can be a complicated function of the local medium density and the beam intensity, leading to non-linear propagation of light. Here we focus on the imaginary part of $\chi$ which is responsible for the absorption of the probe field by the atomic cloud. To solve equation~\eqref{eq:diff_light_prop}, one may take into account the local intensity $I(x)$ to estimate $\chi(x)$. Nevertheless, here we will assume constant light intensity experienced by the atoms resulting in $\chi\big(I,n(x)\big)$, neglecting non-linear light propagation effects. This approximation is justified since we observe the cloud under transparency conditions where $\left|\chi\right| \ll 1$, but small deviations might be observed for large absorption levels \cite{Hofmann2013}. Considering a Gaussian profile for the density $n(x)=n_0 \exp{(-x^2/2\sigma_x^2)}$, we can numerically solve equation~\eqref{eq:diff_light_prop} to calculate the intensity along the propagation through the cloud. As an example, figure~\ref{fig:LightPropNoiseSNR}a shows the light intensity $I(x)$ in the three considered cases, the two-level one using $\chi_{2l}^{~}$ from eq.~\eqref{eq:chi_2lvl}, the response under EIT conditions using $\chi_{eit}^{~}$ from eq.~\eqref{eq:chi_eit}, and the response in the presence of impurities using $\chi_{imp}^{~}$ from eq.~\eqref{eq:chi_IMP}. Experimentally we access the total absorption $A=(I_{in}-I_{trans})/I_{in}$ given by
\begin{equation}
	\label{eq:Absorption}
	A = 1-e^{-\OD} \text{\hspace{3mm} with \hspace{3mm}} \OD=\int_{-\infty}^{+\infty} k \cdot \text{Im} \left[ \chi \Big(n(x),I \Big) \right] dx
\end{equation}
where $\OD$ is the optical density.

To collect only information about the impurities embedded in the atomic gas, we compare the transmitted light under EIT without impurities $I_{eit}$, and with impurities $I_{imp}$. Their presence can therefore be revealed through an additional absorption $A_{add}=(I_{eit}-I_{imp})/I_{eit}$
\begin{equation}
	\label{eq:effective_absorption}	
	A_{add} = 1- e^{-\Delta \OD} \text{\hspace{3mm} with \hspace{3mm}} \Delta \OD = \OD_{imp} - \OD_{eit}
\end{equation}
that we can experimentally access. This additional absorption represents the \emph{signal} emerging from the impurities. It changes from 0 when transparency is not affected by their presence to $A_{2l}$ when impurity-probe interactions fully break the transparency. To get more insight into this quantity it is convenient to rewrite it using eqs.~\eqref{eq:chi_eit} and~\eqref{eq:chi_IMP} into
\begin{equation}
	\label{eq:signal_simplified}
	\Delta \OD= f_{ir}  \int_{-\infty}^{+\infty} k (1-f_{rr}) \text{Im} \left[ \chi_{2l}^{~} - \chi_{eit}^0 \right] dx.
\end{equation}
Under our assumptions only the density is spatially-dependent, but it comes into play in the expressions of the susceptibilities $\chi_{2l}^{~}$ and $\chi_{eit}^0$ as well as in the fraction $f_{rr}$. Measuring the additional absorption $A_{add}$ allows to determine the impurity density $n_i$ which is included in $f_{ir}$, under the assumption of knowing the other parameters involved in eq.~\eqref{eq:signal_simplified}. In order to physically interpret this equation, a first consideration can be done at low probe-probe interactions when $f_{rr} \to 0$: the presence of the impurities blockades a volume fraction $f_{ir}$ in which each probe atom will become an additional absorber leading to a change of susceptibility from $\chi_{eit}^0$ to $\chi_{2l}^{~}$. Then $\Delta \OD$ would be the optical density of these additional absorbers. In the general case, some of these probe atoms would already behave as additional absorbers due to the probe-probe interactions, leading to a reduction of the signal by a factor $1-f_{rr}$. 

Equation~\eqref{eq:signal_simplified} contains all the information needed to optimize the detection of impurities by IEI. The ideal EIT susceptibility $\chi_{eit}^0$ should be minimized by reducing the effective decay rate $\gamma_{gr}$, for instance by decreasing the laser linewidths. The two-level susceptibility $\chi_{2l}^{~}$ should be maximized by avoiding any saturation effect. Additionally the fraction $f_{rr}$ must be kept small compared to 1 such that probe-probe interactions do not affect the transparency too much. This can be done by increasing the EIT bandwidth $\sigma_{eit}$ using large coupling Rabi frequency $\Omega_c$ or by reducing the non-interacting Rydberg population $\rho_{rr}^0$. In the weak probe regime the latter is equal to $\Omega_p^2/\Omega_c^2$ and can therefore be minimized using low probe intensities. Finally, it is obviously needed that the impurities blockade a large fraction of the volume ($f_{ir}\rightarrow 1$). At a given density of impurities, the EIT bandwidth must be smaller than the impurity-probe interaction strength $C_3^{ir}$, in order to have a large blockade radius $R_{ir}$.

\begin{figure}[t]
\centering
\includegraphics[width=\linewidth,keepaspectratio=true]{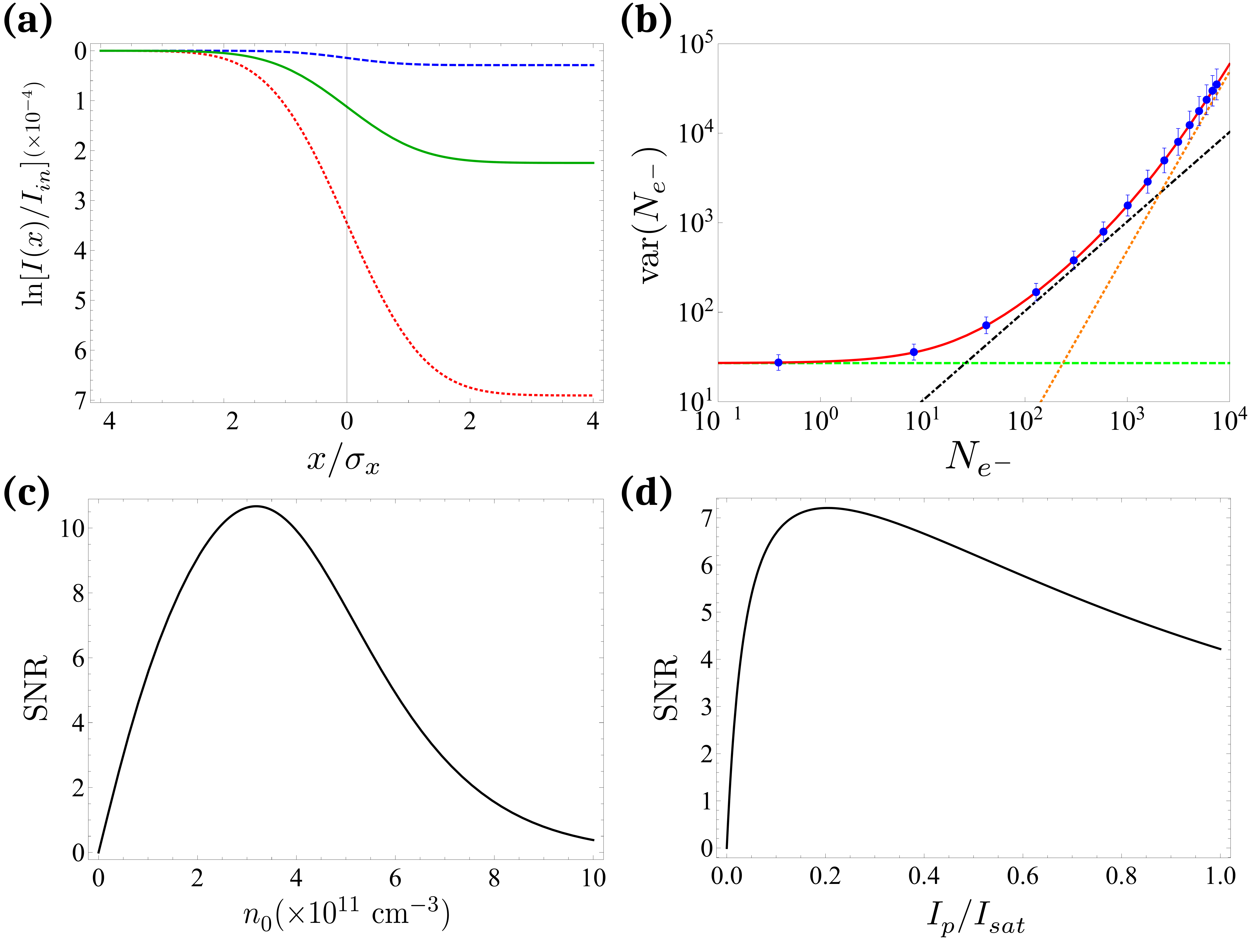}
\caption{Signal and noise estimation from the model. {\bf(a)} Simulation of light propagation through a one-dimensional Gaussian cloud of atoms of width $\sigma_x$ as a function of the rescaled distance $x/\sigma_x$. The red dotted line describes the two level system response with $\Omega_p=2\pi \cdot 1\,\mega\hertz$ and $n_0=1\cdot10^{11}\,\centi\meter^{-3}$, while the blue dashed line refers to the one under EIT ($\Omega_c=2\pi \cdot 10\,\mega\hertz$). The stray absorption is induced by probe-probe interactions estimated here with $C_6^{rr}=2\pi \cdot 1.8\,\rm{GHz}\cdot\rm{\mu m^6}$. The green solid line represents the case of EIT with an impurity density of $n_i=5\cdot10^8\,\centi\meter^{-3}$, assuming an interaction strength $C_3^{ir}=2\pi \cdot 1.7\,\rm{GHz}\cdot\rm{\mu m^3}$. {\bf(b)}~Variance $\text{var}(N_{e^-})$ vs. mean value $N_{e^-}$ of the electron number on our camera (blue points). The mean values and its variances are calculated for each pixel over 50 repetitions and averaged over the exposed area, then fitted with a second order polynomial (eq. \eqref{eq:VarNel}) whose parameters are in agreement with the expected ones. At low signal the noise is dominated by the readout noise of the camera (green dotted line, $r_N^2=27.0 \pm 0.1$), at intermediate levels by the photon shot noise (black dash-dotted line, $a=1.031\pm0.004$), while above $10^3 \rm{e^-/px}$ the noise scales quadratically with the signal (orange dotted line, $b=(4.90\pm0.02)\cdot 10^{-4}$). {\bf(c)} and {\bf(d)} Estimated signal to noise ratio as a function of the peak density of the atomic cloud $n_0$ and of the probe intensity $I_p/I_{sat}$, at a given exposure time $t_{exp}=10\,\rm{\mu s}$ and with the same parameters than (a). The noise is estimated from (b) and assuming $10\%$ of density fluctuation.}
\label{fig:LightPropNoiseSNR}       
\end{figure}

To determine the sensitivity of IEI, we compare the signal given by $\Delta \OD$ to the noise present in the measurements. The first source of noise arises from the detection process of light on the CCD camera. The latter converts the incoming photons on each pixel into electrons with a quantum efficiency $Q_e$ such that the number of electrons per pixel $N_{e^-}$ is proportional to $It_{exp}Q_e$, with $t_{exp}$ the exposure time of the imaging. The detection of these electrons is affected by three main noise sources: electronic noise $r_N$ introduced by the charge readout (constant at each chosen camera setting), photon shot noise and noise originating from other defects of the imaging and detection system which is proportional to the intensity. We can then express the expected noise level for a given number of electrons $N_{e^-}$ by its variance
\begin{equation}
  \label{eq:VarNel}
  \text{var}(N_{e^-}) = r_N^2 + N_{e^-} + a N_{e^-}^2
\end{equation}
where $a$ is a fitted constant to take into account the intensity noise of the detection. These constants have been measured for our imaging system by fitting a second order polynomial to the variance of the measured number of electrons on the camera for different light intensities and the extracted parameters are in very good agreement with the specified ones (fig.~\ref{fig:LightPropNoiseSNR}b). The lowest impact of noise on the signal is in the photon shot-noise limited regime (black dash-dotted line). Each measurement of transmitted light on the CCD camera will lead to a statistical error, finally resulting in a noise of variance var$_I(A_{add})$ for the additional absorption.

The second main source of noise are atomic density fluctuations $n$, including atom shot noise and pure fluctuations of the trap loading over repetitions of the experiment. As the signal is obtained by comparing the outgoing light with and without impurities on two different repetitions, one must take this noise source into account through its variance var$_{n}(A_{add})$. One can thus define a signal-to-noise ratio SNR as
\begin{equation}
	\label{eq:SNR}
	\text{SNR}= \frac{A_{add}}{\sqrt{\text{var}_I(A_{add})+\text{var}_n(A_{add})}}
\end{equation}
that has to be larger than 1 to be able to detect impurities in the gas.

Equation~\eqref{eq:SNR} is used to estimate the SNR we can realistically expect in the experiment and to optimize the experimental settings that would lead to enhanced sensitivity, each of them having multiple influences on the SNR. In figure~\ref{fig:LightPropNoiseSNR}c,d we plot the SNR as function of the experimental parameters $n_0$ and $I_p$, respectively. The first graph illustrates the loss of signal at low density, due to the absence of additional absorbers, and the reduction of sensitivity at high density, where probe-probe interaction induced absorption is so large that the detected signal is weak and dominated by noise. At intermediate density an optimum regime is reached. The second graph shows that noise is predominant at low probe intensity, while the additional absorption gets smaller at large intensity due to increasing probe-probe interactions. In these simulations the predicted SNR is much larger than 1, therefore we can rely on IEI to detect impurities embedded in a cold atom gas.

\section{Experimental setup}
\label{sec:setup}

We perform IEI experiments using the experimental setup described below. We first introduce the preparation of the ultracold atomic cloud using optical dipole traps, then we present the atomic state preparation procedure as well as the detection methods available. 

\subsection{Preparation of an atomic sample in optical dipole traps}
\label{subsec:SetupTightTrap}

In order to perform IEI experiments, we first apply common laser cooling and trapping techniques~\cite{Ketterle1999} to create an ultracold gas of $^{87}$Rb Rydberg atoms as described in detail in~\cite{Hofmann2014}. The atoms are pre-cooled in a 2D-MOT, then loaded into a 3D-MOT in the science chamber~\cite{Dieckmann1998,Schoser2002} and directly transferred into a "reservoir" optical dipole trap (ODT). This trap is composed of two weakly-focused laser beams crossing at a small angle and generated by a $50\,\rm{W}$ single frequency fiber amplifier laser at a wavelength of $1064\,\rm{nm}$, resulting in an elongated cigar-shaped atomic cloud of width of $\sim 40\,\rm{\mu m}\times 40\,\rm{\mu m} \times 800\,\rm{\mu m}$ at $1/e^2$, tilted from the $x$ probing direction by $45\degree$ (fig.~\ref{fig:setup}). This geometry allows for efficient loading from the 3D-MOT leading to large densities up to $5\cdot 10^{11}\,\centi\meter^{-3}$ with typical temperatures of $40\,\micro\kelvin$.

\begin{figure}[ht]
\centering
\includegraphics[width=\linewidth,keepaspectratio=true]{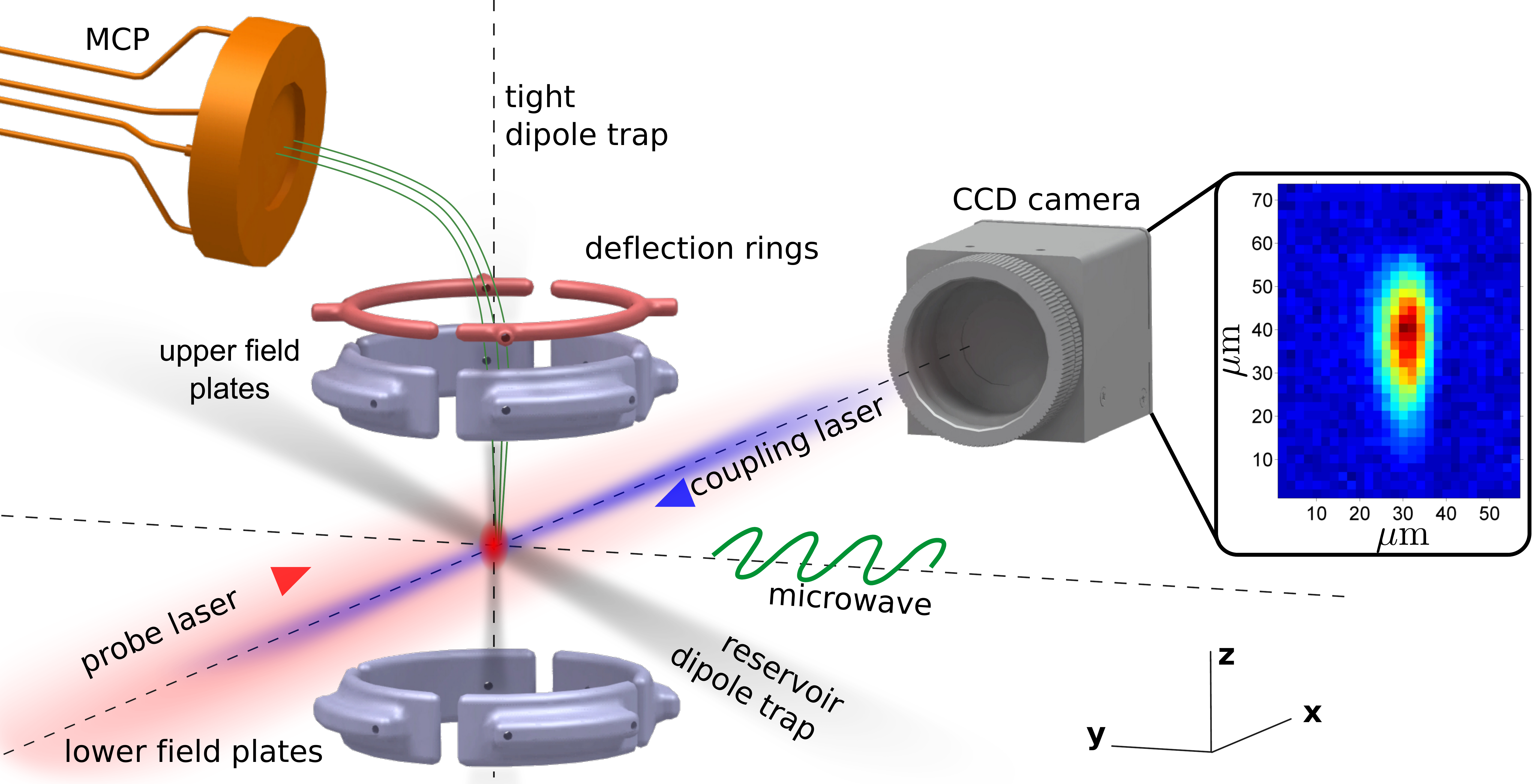}
\caption{Experimental setup for excitation and detection of Rydberg atoms in an ultracold $^{87}\rm{Rb}$ gas. Pre-cooled atoms are transferred from a magneto-optical trap into the reservoir optical dipole trap. An additional tight dipole trap in the perpendicular direction is used to create small and dense samples. After being released from the trap, the atoms are excited to Rydberg states by a combination of a $780\,\rm{nm}$ probe laser, a counter-propagating $480\,\rm{nm}$ coupling laser and microwave radiation. An electrode structure composed of 8 field plates is used for precise electric field control and for field ionization of Rydberg atoms, whereas two deflection rings guide the ions (green trajectories) onto a micro-channel plate (MCP) detector. Complementary information on the spatial distribution of ground state atoms is acquired in parallel by absorption imaging with a CCD camera. The inset shows the two-level absorption of the atoms in the tight trap (in false color).}
\label{fig:setup}
\end{figure} 

Experiments realized in a trap of similar geometry have shown the emergence of diffusive transport effects~\cite{Guenter2013}, since the Rydberg impurities and the probe atoms are liable to resonant dipolar exchange. To circumvent such effects, the atoms can be confined into a small region comparable to the typical Rydberg blockade volume using an additional vertical tight ODT. We temporarily transfer the atoms from the reservoir trap to a tightly focused third "dimple" ODT aligned on the $y$ direction. All-optical evaporative cooling is used to increase the phase-space density of the atoms whilst decreasing their temperature~\cite{Clement2009,Grimm2000}. Then the tight ODT is loaded by increasing its potential depth while reducing the one of the dimple trap. The final cloud contains up to 1500 atoms in a volume of $\sim 8\,\rm{\mu m}\times 22\,\rm{\mu m} \times 8\,\rm{\mu m}$ (width at $1/e^2$), providing densities up to $5\cdot 10^{11}\,\centi\meter^{-3}$ as well as low temperatures of $\sim 1\,\micro\kelvin$. Therefore only very few impurities can be excited in this tight trap before reaching the fully-blockaded regime, where the exchange dynamic is inhibited.


\subsection{Excitation and detection techniques}
\label{subsec:SetupExcDet}

After completion of the loading of the atoms into one of the two dipole traps, a static magnetic field of $6\,\gauss$ is applied along the $x$ direction to define the quantization axis. Initially both the $F=1$ and $F=2$ manifolds of the $^{87}$Rb ground state $\ket{5S_{1/2}}$ are populated. To prepare a clean initial ground state we first remove the atoms in the $F=2$ manifold by switching off the repumping light at the end of the MOT cooling stage. Then we transfer populations from the state $\ket{5S_{1/2},F=1,m_F=1}$ to the selected ground state $\ket{g}=\ket{5S_{1/2},F=2,m_F=2}$ by means of a microwave Landau-Zener adiabatic sweep. Tuning the duration of the sweep allows for controlling the ground-state atomic density of the sample. In order to optically image the ultracold gas we resonantly couple a $780\,\rm{nm}$ weak probe laser beam to the $\ket{g}\leftrightarrow\ket{e}=\ket{5P_{3/2},F=3,m_F=3}$ transition and collect the transmitted light onto an Andor iXon Ultra CCD camera (fig.~\ref{fig:setup}) via a nearly diffraction limited imaging system with a resolution of $4.8\,\micro\meter$ (Rayleigh criterion). An additional counter-propagating $480\,\rm{nm}$ strong coupling beam is focused on the center of the cloud to couple atoms to a $\ket{nS}$ Rydberg state and provide conditions for EIT. 

To minimize the absorption under EIT coupling it is important to reduce the frequency fluctuations of the lasers which effectively contribute to the dephasing rate $\gamma_{gr}$, affecting the transparency of the atoms (see eq.~\eqref{eq:chi_EIT0}). In our setup the probing and coupling lasers are both frequency stabilized to a passive high finesse ultra-low-expansion glass Fabry-P\'{e}rot cavity~\cite{Gregory2015,Aikawa2011} via the Pound-Drever-Hall method (PDH)~\cite{Black2001}. The cavity mirrors are dual-wavelength coated, allowing to stabilize both lasers at the same time. 
To generate the PDH error signal and to have a broad frequency tunability range we pass each beam through a broadband fibre-coupled electro-optic modulator to which we apply, by means of a power combiner, a modulation and an offset frequencies. We estimate the frequency stability of our lasers by measuring the root mean square instantaneous frequency deviation relative to a cavity mode of a reference active Fabry-P\'{e}rot cavity (Sirah Eagle Eye). For timescales longer than $3\,\micro\second$ we measure a linewidth below $10\,\kilo\hertz$ which allows perfect transparency in absence of interactions ($\chi_{eit}^0 \approx 0$).

To prepare the Rydberg $P$ impurities we perform a three-photon excitation. The first stage of the excitation is done using another $780\,\nano\meter$ laser beam propagating along the vertical direction (not shown in fig.~\ref{fig:setup} for clarity) and linearly polarized such that it generates both $\sigma^+$ and $\sigma^-$ polarization, only the $\sigma^+$ one being relevant for the excitation. The $\sigma^-$-polarized light leads to off-resonant processes that can be neglected in good approximation. For the second stage we use the same blue beam that is involved in EIT. Additionally we employ an Anritsu MG3697C microwave synthesizer which allows to address transitions between Rydberg states with different angular momentum, e.g. $\ket{nS}\leftrightarrow \ket{nP}$ transitions. The microwave radiation is emitted in the science chamber by a simple antenna without any control of the polarization.

Finally, the experimental apparatus has an electrode structure which allows to apply and control the electric fields in the three dimensions, which is necessary for Rydberg atoms due to their high atomic polarizability~\cite{Low2012,Sassmannshausen2013,OSullivan1985}. 
Moderate electric fields can be applied to DC-Stark shift the energy levels of the Rydberg atoms or to tune their F\"{o}rster defect and therefore the strength of the interactions~\cite{Vogt2006,Westermann2006,Ryabtsev2010,Nipper2012,Gurian2012}. The electrodes also enable us to count the number of Rydberg atoms present in the sample by ionizing them and guiding the generated ions to a micro-channel plate (MCP) particle detector with an estimated overall detection efficiency of $\eta\approx0.4$. 

\section{Off-resonant excitation of Rydberg P states}
\label{sec:Excitation}

To prepare $nP$ Rydberg states of $^{87}$Rb a two-photon excitation is prohibited by the selection rules. Instead we use a three-photon excitation making use of two optical photons and an additional microwave photon. This approach allows us to simultaneously excite $nS$ and $nP$ states, both required for the experiments presented in this work.

The excitation scheme is presented in figure~\ref{fig:two-level-approx}a. The $\sigma^+$-polarization of a $780\,\nano\meter$ laser beam couples the ground state $\ket{g}$ to the intermediate state $\ket{e}$.  Hence, a $480\,\nano\meter$ $\sigma^-$-polarized laser couples $\ket{e}$ to the Rydberg state $\ket{r}=\ket{nS_{1/2},m_j=1/2}$ with $n=42$. The third stage of the excitation to the impurity state $\ket{i}=\ket{nP_{3/2},m_j=3/2}$ is done using a microwave photon with a frequency around $53.9\,\giga\hertz$. This four-level system is then described by the Hamiltonian
\begin{equation}
	\label{eq:hamiltonian}
	\begin{split}
	\hat H =& - \Delta_e \ket{e}\bra{e} - \Delta_r \ket{r}\bra{r} - \Delta_{mw} \ket{i}\bra{i}\\
	 &+ \left( \frac{\Omega_e}{2} \ket{e}\bra{g} + \frac{\Omega_r}{2} \ket{r}\bra{e} + \frac{\Omega_{mw}}{2} \ket{i}\bra{r} + h.c. \right)
	\end{split}
\end{equation}
with $\hbar = 1$, $\Delta_e,\,\Delta_r,\,\Delta_{mw}$ the one-photon, two-photon and three-photon detunings respectively, and $\Omega_e,\,\Omega_r,\,\Omega_{mw}$ the Rabi frequencies of each transition.

\begin{figure}[t]
\centering
\includegraphics[width=\linewidth,keepaspectratio=true]{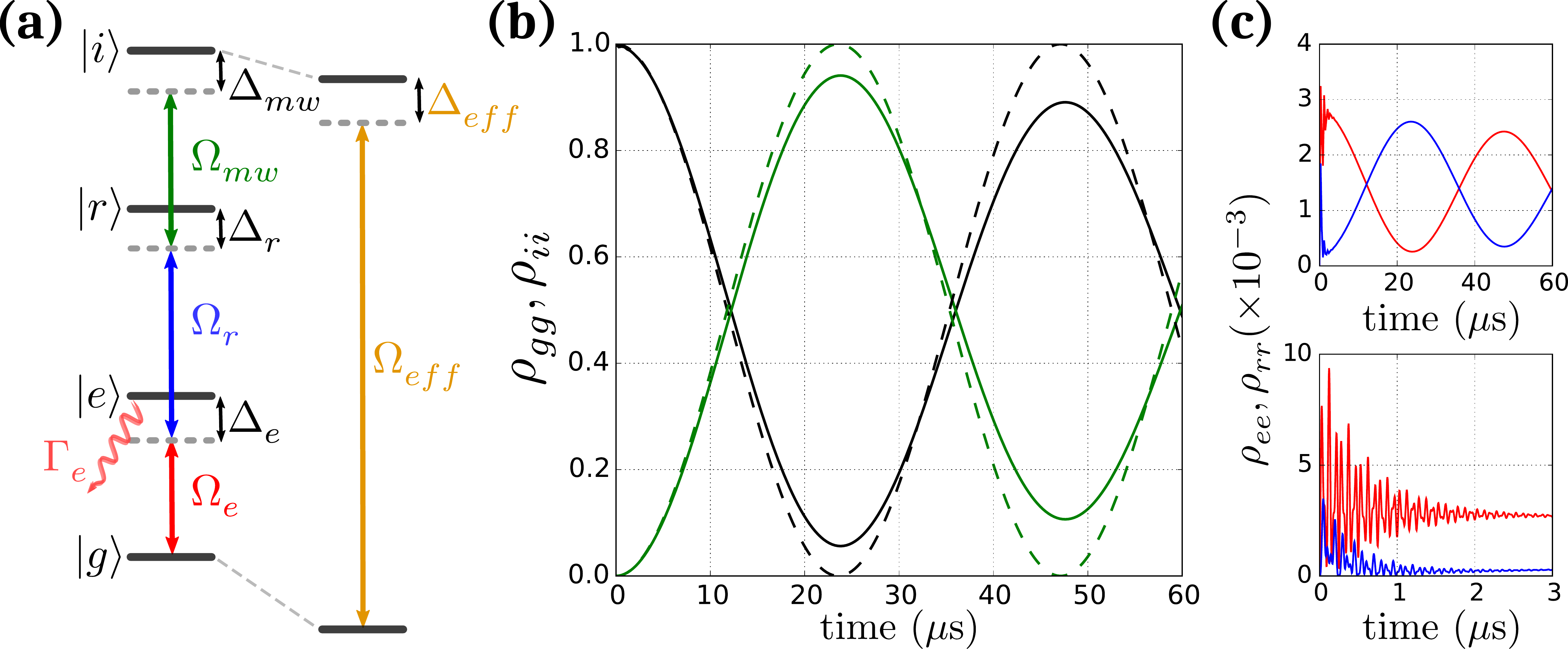}
\caption{Excitation of Rydberg $P$ states. {\bf(a)} Three-photon excitation scheme. 
The ground state $\ket{g}$ is coupled to the impurity state $\ket{i}$ by two optical fields and one microwave field with Rabi frequencies $\Omega_e$, $\Omega_r$ and $\Omega_{mw}$, respectively. Large intermediate detunings $\Delta_e$ and $\Delta_r$ prevent extensive populations of the two intermediate states $\ket{e}$ and $\ket{r}$. 
{\bf(b)} Simulation of the time-evolution of the state populations $\rho_{gg}$ and $\rho_{ii}$ (resp. black and green lines), assuming the initial population in the ground state, with $\Delta_e=\Delta_r=- 2\pi\cdot 100\,\rm{MHz}$, $\Omega_e=\Omega_{mw}=2\pi\cdot 2.4\,\rm{MHz}$ and $\Omega_r=2\pi\cdot25\,\rm{MHz}$. The exact calculation is compared to the effective two-level model (dashed lines). The observed difference is due to the absence of the decay $\Gamma_e$ in the later. 
{\bf(c)} Populations $\rho_{ee}$ and $\rho_{rr}$ simulated in the same regime. With a maximum fraction of $3\cdot 10^{-3}$ these states are weakly populated. We observe on short timescales very fast oscillations damped after $2\,\micro\second$, justifying the adiabatic elimination, then the populations follow a similar evolution to $\ket{g}$ and $\ket{r}$.}
\label{fig:two-level-approx}       
\end{figure}

\subsection{Effective two-level approximation of the four-level system}

To avoid populating the two intermediate states we use large intermediate detunings $\Delta_e$ and $\Delta_r$ compared to the Rabi frequencies $\Omega_e$, $\Omega_r$ and $\Omega_{mw}$. In this regime, the four-level excitation can be described by an effective two-level system that we theoretically develop in the following. The model is very helpful to derive the best strategy to efficiently excite Rydberg $P$ states without any undesired additional effects.

To simulate the three-photon excitation one can use the optical Bloch equations in the four-level basis and numerically solve them for a set of the six experimental parameters given by eq. \eqref{eq:hamiltonian}. However, the numerical simulations do not give insight into the excitation process to find the optimal settings for efficient impurity preparation. Since the excitation is done in the off-resonant regime $\Delta_e,\Delta_r~\gg~\Omega_e,\Omega_r,\Omega_{mw}$, the evolution of the intermediate state populations exhibit two very different timescales, leading to a rapidly oscillating term at a frequency $\Delta_{e,r}$ and a slow-varying envelope that evolves together with the states $\ket{g}$ and $\ket{i}$. After a short-time, the fast oscillations average to zero due to the decay of the short-lived intermediate state $\ket{e}$. The time derivative of the $\ket{e}$ and $\ket{r}$ populations can then be set to zero as they are dominated by the fast-oscillating terms. By introducing the steady state solutions in the optical Bloch equations, one can eliminate all the terms involving $\ket{e}$ and $\ket{r}$ in order to write an effective equation valid in the limit of long timescales, which directly couples $\ket{g}$ to $\ket{i}$. This approximation is known as adiabatic elimination~\cite{Linskens1996,Brion2007}. With $c_x$ the projection of the wavefunction to the state $\ket{x}$ ($x=g,e,r,i$), one obtains the following coupled equations
\begin{eqnarray}
	\label{eq:Schrodinger-two-level}
	\dot{c}_g &=& -i S_g c_g + i \frac{\Omega_{\eff}}{2} c_i \\
	\dot{c}_i &=& -i (\Delta_{mw} - S_i) c_i + i \frac{\Omega_{\eff}}{2} c_g
\end{eqnarray}
where $S_g$ and $S_i$ are the AC-Stark shifts of the states $\ket{g}$ and $\ket{i}$ defined by
\begin{equation}
	\label{eq:AC-Stark-shift}
	S_g = \dfrac{\Omega_e^2}{4\Delta_e - \frac{\Omega_r^2}{\Delta_r}} \hspace{2cm} S_i = \dfrac{\Omega_{mw}^2}{4\Delta_r - \frac{\Omega_r^2}{\Delta_e}}
\end{equation}
and $\Omega_{\eff}$ is the effective Rabi frequency of the four-level system defined by
\begin{equation}
	\label{eq:eff-Rabi}
	\Omega_{\eff} = \dfrac{\Omega_e \Omega_r \Omega_{mw}}{4 \Delta_e \Delta_r - \Omega_r^2}.
\end{equation}
Equation~(\ref{eq:AC-Stark-shift}) simply shows that one can reduce the effect of the three off-resonant fields to AC-Stark shifts of the two outer states $\ket{g}$ and $\ket{i}$ and to an effective Rabi frequency $\Omega_{\eff}$ coupling these states. The AC-Stark shifts are calculable in a perturbative approach. By setting the ground-state energy to zero via a rotation in the rotating frame, the differential equations can be rewritten as
\begin{eqnarray}
	\label{eq:eff-two-level}
	\dot{c}_g &=& i \frac{\Omega_{\eff}}{2} c_p\\
	\dot{c}_i &=& -i \Delta_{\eff} c_i + i \frac{\Omega_{\eff}}{2} c_g.
\end{eqnarray}
with $\Delta_{\eff}=\Delta_i +S_g -S_i$ the effective detuning. At this stage one can identify the Schr{\"{o}}dinger equation of a two-level system with Rabi frequency $\Omega_{\eff}$ and detuning $\Delta_{\eff}$. Therefore the whole system can be considered as an effective two-level system where conditions for the inversion of state populations are known.

This effective description is illustrated in figure~\ref{fig:two-level-approx}b, where the populations of the four levels are plotted as a function of the excitation time and compared to full numerical solutions of the optical Bloch equations. It clearly shows that the two intermediate states are negligibly populated and that the two outer states behave like a two-level system, exhibiting an evolution similar to two-level Rabi oscillations. We notice that the effective model does not predict the decay of the oscillations which would be a consequence of the decay of the intermediate state, not considered in the current effective model. To take it into account we use an effective operator formalism for open quantum systems to derive an effective master equation~\cite{Reiter2012}. It results that the decay $\Gamma_e$ effectively leads to various terms in the two-level model which can, on long timescales, either kill or enhance coherences. In the following discussion we neglect this decay since it has no crucial impact on the regime we are working on.

Using this approach we can derive an optimized excitation scheme of the Rydberg state $\ket{i}$. Due to the AC-Stark shifts, the intuitive resonance condition with $\Delta_{mw}=0$ is not valid. From the effective two-level model we deduce a modified resonance condition that must be fulfilled to efficiently transfer an atom from $\ket{g}$ to $\ket{i}$
\begin{equation}
	\label{eq:eff-resonance}
	\Delta_{\eff} = \Delta_i + \dfrac{\Omega_e^2}{4\Delta_e - \frac{\Omega_r^2}{\Delta_r}} - \dfrac{\Omega_{mw}^2}{4\Delta_r - \frac{\Omega_r^2}{\Delta_e}} = 0.
\end{equation}
In addition, two other constraints must be considered before performing the experiment. On one hand we aim to excite impurities on a timescale shorter than the impurity lifetime, typically $\tau_i\approx50\,\micro\second$, leading to the condition
\begin{equation}
	\label{eq:condition-omega-eff}
	\Omega_{\eff} \leq \tau_i^{-1}.
\end{equation}
On the other hand we want to avoid populations in both intermediate states. Populations in the Rydberg state $\ket{r}$ cannot be resolved by our field ionization setup and would be mistaken for impurities, while population in $\ket{e}$ could lead to scattering
and heating of the cloud. In the far-detuned regime, $c_e$ and $c_r$ are very accurately approximated by $\Omega_e/\Delta_e\cdot c_g$ and $\Omega_{mw}/\Delta_{mw}\cdot c_i$, respectively. To satisfy both conditions, it is favourable to keep $\Omega_e$ and $\Omega_{mw}$ relatively small, while $\Omega_r$ can remain large as long as it satisfies the far-detuned condition $\Omega_r \ll \Delta_e, \Delta_r$. Our choice of states, involving the same principal quantum number $n$ for both the probe and impurity Rydberg states, allows to use the same $480\,\nano\meter$ laser field for the second stage of impurity excitation and for the EIT coupling. Hence $\Delta_r=0$ and the detunings must fulfil $\Delta_e=\Delta_r$. In all experiments we set them to $\sim 2\pi \cdot 100\,\mega\hertz$. In these conditions the red beam heating effect is negligible if the Rabi frequency $\Omega_e$ is below a critical value of typically $2\pi \cdot 2.5\,\mega\hertz$. Hence, $\Omega_r$ and $\Omega_{mw}$ can be adjusted to respect the above-mentioned conditions. The simulations in fig.~\ref{fig:two-level-approx} show an example of efficient excitation with populations in the intermediate states smaller than $3\%$.

\subsection{Three-photon spectroscopy of 42P Rydberg states}
\label{sec:experimental_excitation}

\begin{figure}[b]
\centering
\includegraphics[width=\linewidth,keepaspectratio=true]{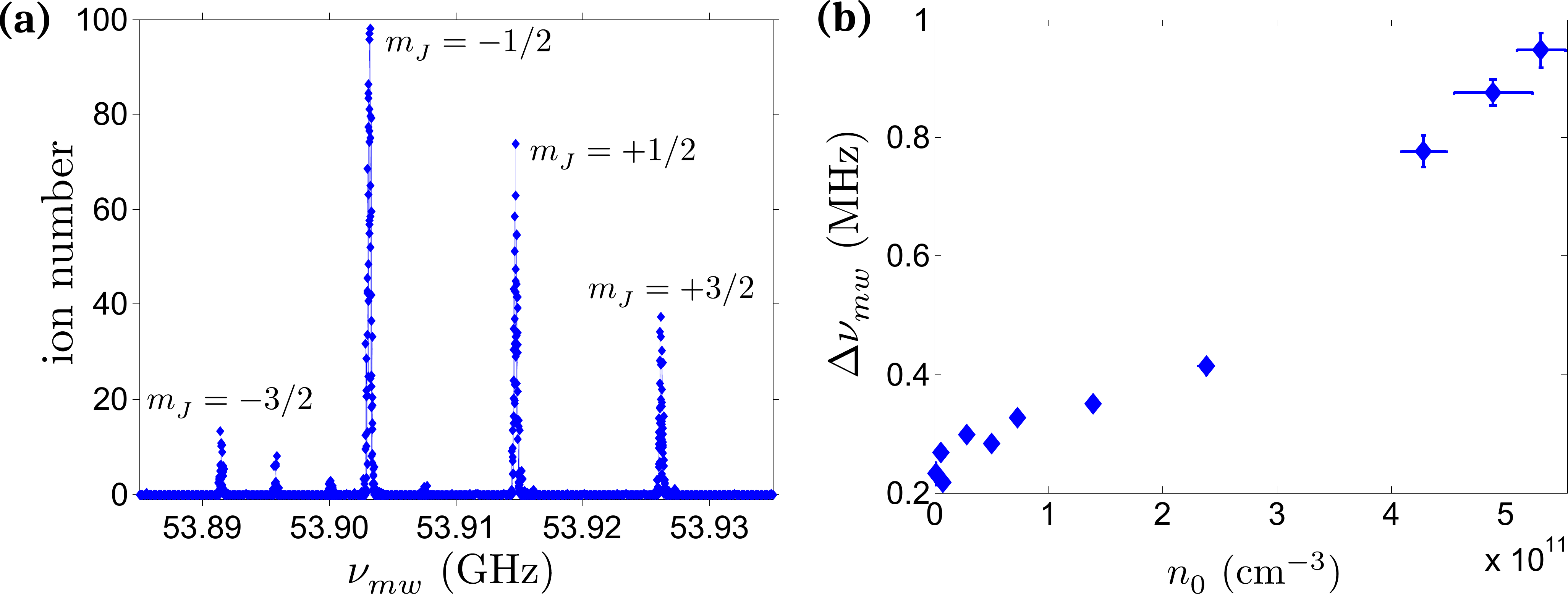}
\caption{Three-photon spectroscopy of the $\ket{42P_{3/2}}$ states. {\bf(a)} Microwave frequency scan around the three-photon resonance, at a peak density $n_0=(3.30\pm0.18)\cdot10^{9}\,\rm{cm}^{-3}$ where interaction effects do not play a significant role. The atoms, initially prepared in $\ket{g}$, are off-resonantly driven to $\ket{e}$ and $\ket{r}=\ket{42S_{1/2},m_j=+1/2}$. A detuned microwave radiation finally excites the impurity state $\ket{i}$ by compensating the energy mismatch. Only the Zeeman substates $m_j=+3/2,+1/2,-1/2$ are addressed with $\sigma^{+},\pi,\sigma^{-}$ polarization components, respectively. The Rabi frequencies are calibrated to be $\Omega_e=2\pi\cdot9.7\,\rm{MHz}$, $\Omega_r=2\pi\cdot25\,\rm{MHz}$, $\Omega_{mw}=2\pi\cdot12\,\rm{MHz}$. The spectrum shows a significant population of $\ket{42P_{3/2},m_j=-3/2}$, coming from a residual $\pi$ polarization component of $\Omega_e$. The additional residual peaks come from a cycling process that populates different $m_F$ sub-states of $\ket{5S_{1/2},F=2}$. A magnetic field $B_x=6.43\,\rm{G}$ is estimated from the fine splitting observed between the $m_j$ peaks. 
{\bf(b)} Density dependence of the width of the $\ket{42P_{3/2},m_j=3/2}$ resonance. In contrast to (a), many-body interaction effects play a major role on the lineshape at higher densities. The spectrum is acquired with $\Omega_e=2\pi\cdot5.6\,\rm{MHz}$, $\Omega_r=2\pi\cdot25\,\rm{MHz}$, $\Omega_{mw}=2\pi\cdot12\,\rm{MHz}$.}
\label{fig:spectrum_low}       
\end{figure}

In order to demonstrate the detection of $nP$ Rydberg states using IEI, we choose the states $\ket{r}=42S_{1/2}$ and $\ket{i}=42P_{3/2}$ as probe and impurity Rydberg states, respectively. 
As previously discussed, the two optical photons only address the Zeeman substate $\ket{42S_{1/2},m_j=+1/2}$. The Rydberg impurity state $\ket{42P_{3/2}}$ is addressed by a microwave field without control of the polarization, allowing for excitation of the $m_j=-1/2,+1/2,+3/2$ states. The magnetic field induces a Zeeman shift that lifts the degeneracy between these states. Scanning the microwave frequency $\nu_{mw}$ we record the spectrum shown in fig.~\ref{fig:spectrum_low}a using field-ionization detection after the excitation pulse. Large numbers of ions are measured at three expected equally-spaced frequencies. In addition a fourth, smaller peak appears at the expected position of the $m_j=-3/2$ Zeeman substate, which can be explained by misalignment of the polarization of the $780\,\rm{nm}$ excitation beam with the vertical direction, resulting in a residual $\pi$-polarized component for which two additional $\sigma^{-}$ photons allow to address the $m_j=-3/2$ state. We also observe small additional peaks with frequency separations exactly corresponding to the ground state Zeeman splitting. Such resonances are signature of transitions from the other hyperfine levels of the ground state $\ket{5S_{1/2},F=2}$ to the $\ket{42P_{3/2}}$ state. Population in these states cannot be explained by imperfections of the state preparation since we initially excite with high-efficiency the pure $m_F=2$ hyperfine state. However they could arise from coupling to the Rydberg state, for which the purity of the $m_F$ quantum number is not preserved.

\begin{figure}[b]
\centering
\includegraphics[width=\linewidth,keepaspectratio=true]{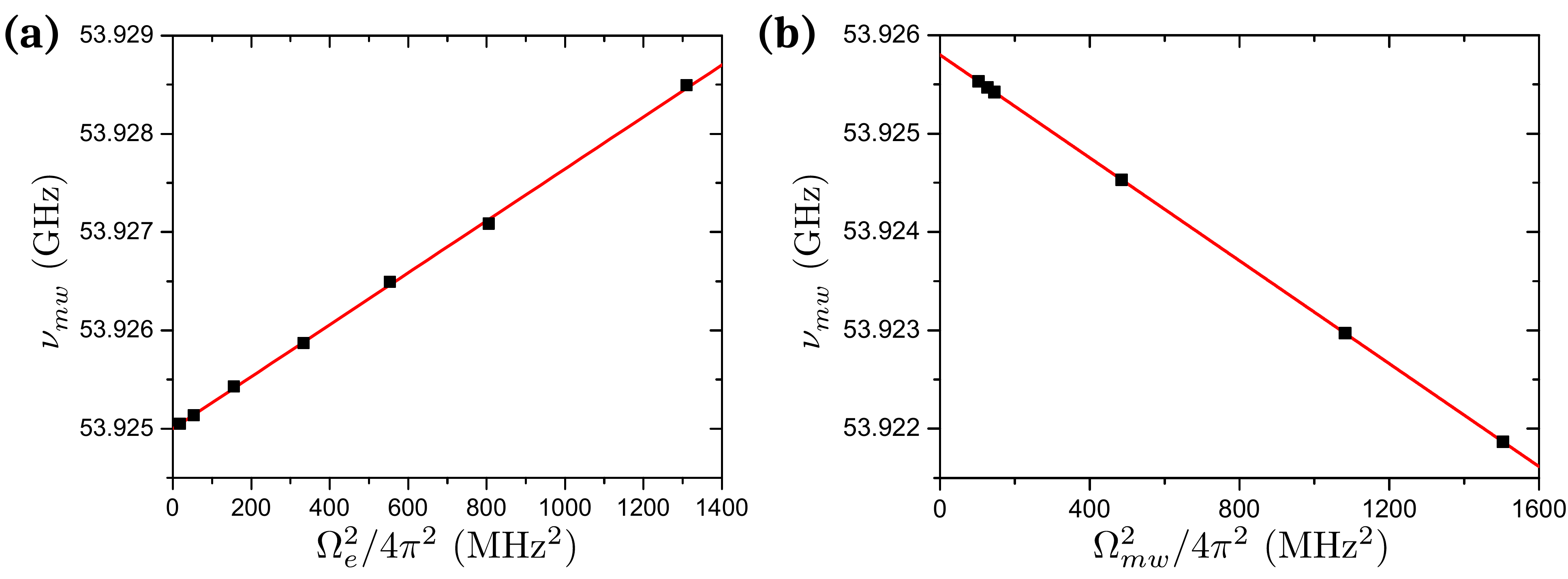}
\caption{Calibration of $\Omega_e$ and $\Omega_{mw}$ from the light shift of the resonance. {\bf(a)} For a given coupling of $\Omega_r$ and $\Omega_{mw}$ the intensity of the red beam $\Omega_e$ is varied. The resonant position of the peak $\ket{42P_{3/2,+3/2}}$ undergoes a quadratic light shift following eq.~\ref{eq:AC-Stark-shift}. {\bf(b)} Analogously, for given $\Omega_e$ and $\Omega_r$, the same procedure can be applied to calibrate $\Omega_{mw}$. We must rely on the fact that the AC-Stark Shift is linear with the power (eq~\ref{eq:AC-Stark-shift}) and therefore find the unperturbed central peak frequency.}
\label{fig:light_shift}
\end{figure} 

The spectrum in fig.~\ref{fig:spectrum_low}a was taken under low-density conditions to minimize multi-particle interaction effects, which have been reported to strongly modify the width of the observed transitions~\cite{Anderson2002,Park2011,Afrousheh2004,Afrousheh2006,Goldschmidt2016}. In our experiments, we perform the excitation of the $\ket{42P_{3/2},m_j=3/2}$ at various densities of ground state atoms and, as shown in fig.~\ref{fig:spectrum_low}b, we observe an increase of the width of the spectral line at higher densities, which constitutes a signature of van der Waals Rydberg-Rydberg interactions.

Three-photon spectroscopy provides a powerful tool to calibrate the Rabi frequencies and the detunings of the fields involved in the excitation process because the resonance condition depends on all these parameters according to eq.~\eqref{eq:eff-resonance}. Only the measurement of the coupling Rabi frequency $\Omega_r=\Omega_c$ is independently realized using a method based on a local fitting of the electromagnetically-induced transparency profile~\cite{Gavryusev2016}. 
By performing three-photon spectroscopy at different $\Omega_e$ (fig.~\ref{fig:light_shift}a) while keeping the other settings constant, we observe that the resonance frequency exhibits a linear behavior with the laser power $P_e \propto \Omega_e^2$ (measured independently), in perfect agreement with eq.~\eqref{eq:eff-resonance}. We determine the unshifted resonance frequency from a linear extrapolation to zero intensity and we can afterwards express the measured resonance frequencies in terms of light shifts $S_g$ of the ground state, allowing us to use equation~\eqref{eq:AC-Stark-shift} to calculate the Rabi frequencies $\Omega_e$. The calibration between $P_e$ and $\Omega_e^2$ is then used to plot the figure~\ref{fig:light_shift}a, revealing a linear relation as would have been expected from the two-level model, thus validating its use.

A similar procedure cannot be applied to calibrate the microwave Rabi frequency $\Omega_{mw}$ since the intensity of the microwave source at the position of the atoms is unknown. Therefore we rely on the same theoretical predictions for the AC-Stark shift, assuming a linear dependence with $\Omega_{mw}^2$. From the variation of the spectral peak position for different microwave intensities, we infer the non-shifted three-photon resonance that leads to a linear relation of the resonance frequency with $\Omega_{mw}^2$ (fig.~\ref{fig:light_shift}b). From it we again calculate the AC-Stark shifts and therefore deduce a calibration of $\Omega_{mw}$. We have checked the consistency of these measurements by measuring the non-shifted transition, setting $\Omega_e=\Omega_{mw}$. We obtain a resonance shifted by $-96\,\rm{MHz}$, in agreement with the experimental calibration of the detuning within $0.5\,\rm{MHz}$. In this way we obtain all the required parameters to perform an efficient and controllable excitation of Rydberg $P$-state impurities.

\subsection{Incoherent excitation of 42P Rydberg impurities}
\label{sec:ImpIncExc}

\begin{figure}[b]
\centering
\includegraphics[width=1.00\linewidth,keepaspectratio=true]{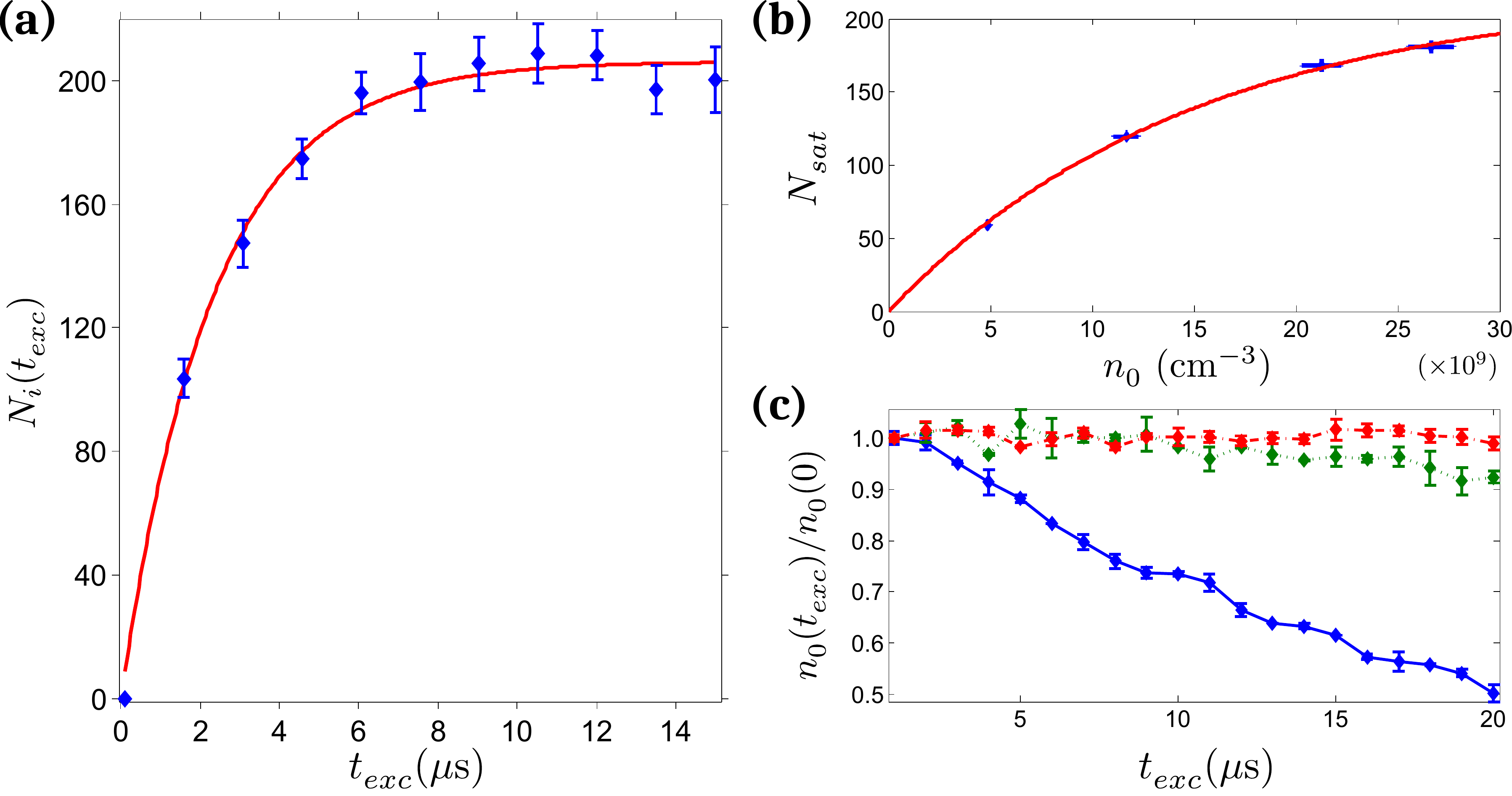}
\caption{Excitation of $\ket{42P_{3/2},m_j={3/2}}$ Rydberg atoms. {\bf(a)} Excitation dynamics as a function of the excitation time $t_{exc}$. The measurement is done after a fixed time-of-flight of $1\,\rm{ms}$, at a peak ground-state density $n_0=(7.26\pm0.21)\cdot10^{9}\,\rm{cm}^{-3}$, with $\Omega_e=\Omega_{mw}=2\pi\cdot10\,\rm{MHz}$, $\Omega_r=2\pi\cdot25\,\rm{MHz}$. The data show a saturation of the number of excitation $N_i$. The saturation is fitted with the function $N_i(t_{exc})=N_{sat}(1-e^{-t_{exc}/\tau})$, obtaining $N_{sat}=206.2\pm2.6$ and $\tau=(2.34\pm0.11)\,\rm{\mu s}$. 
{\bf(b)} Blockade effect on excited impurities. Repeating the same experiment, we observe a saturation of $N_{sat}$ while increasing the ground-state density. To reach high enough densities, the time-of-flight is reduced to $0.5\,\rm{ms}$.
{\bf(c)} Heating induced by scattering on the $\ket{g} \leftrightarrow \ket{e}$ transition. The density is plotted at different excitation times $t_{exc}$ for  $\Omega_e=2\pi \cdot[2.4,\,7.1,\,21]\,\rm{MHz}$ (resp. dash-dotted red, dotted green and solid blue lines). The microwave is switched off to avoid density reduction due to Rydberg excitation. The error bars represent the standard deviation.}
\label{fig:incoherent_exc}       
\end{figure}

Interaction-enhanced imaging experiments require precise control of the number of impurities that we probe. For this purpose it is important to understand the evolution of the impurity population with the excitation time $t_{exc}$. The theoretical model predicts coherent Rabi oscillations between the ground and the impurity states with an effective Rabi frequency $\Omega_{\eff}$ given by eq. (\ref{eq:eff-Rabi}). Yet our measurements show the emergence of a saturation effect after typically $6\,\rm{\mu s}$ (fig.~\ref{fig:incoherent_exc}a), that we attribute to incoherent excitation of the $\ket{42P_{3/2},m_j=3/2}$ impurities Rydberg states. We understand this to be a consequence of a large degree of dephasing introduced by the strong inhomogeneity of the effective Rabi frequency $\Omega_{\eff}$, due to the Gaussian profile of the coupling beam~\cite{Heidemann2007}.

Repeating the experiment for increasing ground-state atomic densities, we observed a similar behavior of the excitation dynamics on the same timescale but with different saturated number of excitations $N_{sat}$. In figure~\ref{fig:incoherent_exc}b this number is plotted versus the ground-state density $n_0$. As can be seen from the plot we observe a second saturation effect that we attribute to the Rydberg blockade between impurities: because of their van der Waals interactions, the total number of impurities in the finite volume of the cloud is limited. Given that the detection efficiency is not known with high accuracy, the value $N_{bl}=218\pm6$ we extract from the exponential fit is consistent with an estimation based on the impurity-impurity blockade radius ($R_{ii}=(C_6^{ii}/\Delta \nu_{mw})^{1/6} \sim 4\,\micro\meter$ with $\Delta \nu_{mw}$ the width of the resonance at low density extracted from fig.~\ref{fig:spectrum_low}b) that would lead to $\sim 230$ impurities in the excitation volume.

The final step to optimize the excitation of impurities consists in minimizing any heating effects due to large excitation Rabi frequency $\Omega_e$, which lead to loss of atoms from the atomic cloud (fig.~\ref{fig:incoherent_exc}c). In order to avoid such effect, in the following we reduce the power of the $780\,\rm{nm}$ laser beam to $\Omega_e=2\pi \cdot 2.4\,\rm{MHz}$ and compensate it by increasing the microwave Rabi frequency $\Omega_{mw}$ to $2\pi \cdot 14\,\rm{MHz}$, such that the effective coupling is not significantly affected.

\section{Imaging of Rydberg 42P atoms}

\label{sec:IEI_experimental}

We discuss in this last section the imaging of Rydberg $\ket{i}=\ket{42P_{3/2},m_j=3/2}$ impurities embedded in a gas of atoms and prepared using the excitation scheme introduced in section~\ref{sec:Excitation}. IEI is performed by coupling the probe atoms to the Rydberg state $\ket{r}=\ket{42S_{1/2},m_j=1/2}$. For the first experimental realization we work in a large volume trap which allows for good statistics in order to validate our model. We then image Rydberg impurities within a small volume to avoid transport effects.

\subsection{Testing the validity of the imaging model}
\label{sec:IEI_test}

To test the validity of the hard-sphere model presented in sec.~\ref{sec:ModelIEI} we first perform an IEI experiment in the large, dense reservoir optical dipole trap (see sec.~\ref{subsec:SetupTightTrap}). The atoms are initially prepared in the ground state $\ket{g}$ at a temperature $T\approx 40\,\rm{\mu K}$. The atomic cloud exhibits a Gaussian profile of width $\sigma_x=41.2\,\micro\meter$ at $1/e^2$ and of tunable density, with peak densities in the center of the cloud $n_0$ up to $1.5\cdot 10^{11}\,\rm{cm^{-3}}$. At a given time, we release the trap and probe the response of the atomic cloud by measuring the absorption in three different configurations: without coupling light nor impurities (two-level response), with the coupling light turned on but no impurities (EIT response), and finally with both the coupling light and the impurities present. To measure the probe light absorption $A$ defined in~\eqref{eq:Absorption}, we record on the CCD camera the probe intensity $I_{trans}$ just after switching off the optical dipole trap as well as the intensity $I_{in}$ after $10\,\milli\second$ when all the atoms have fallen below the light path because of gravity.

\begin{figure}[t]
\centering
\includegraphics[width=\linewidth,keepaspectratio=true]{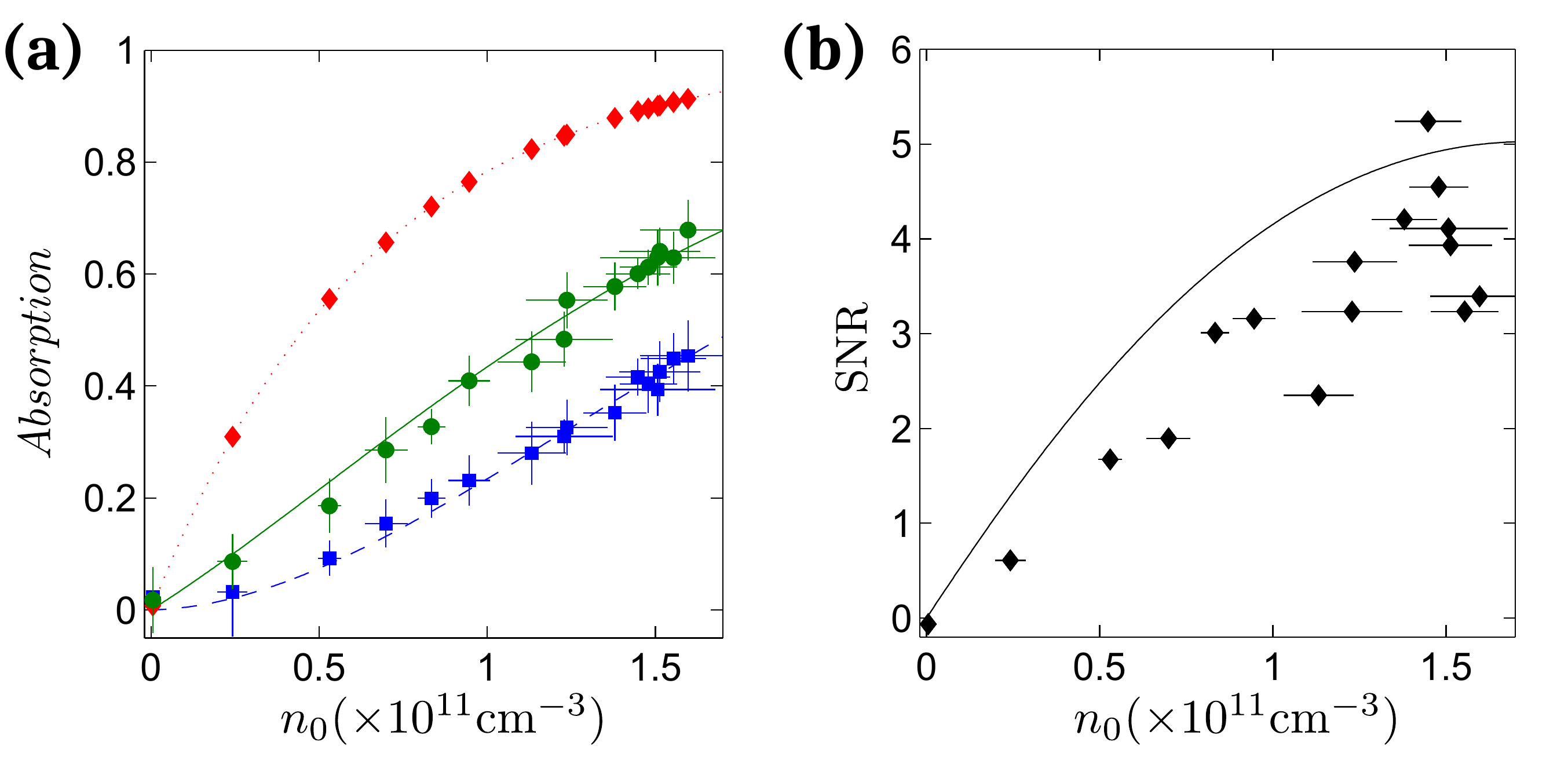}
\caption{Interaction enhanced imaging of $P$-state impurities in a large dipole trap. {\bf(a)} Comparison of the measured and theoretical absorptions $A$ of the probe beam through the atomic cloud, as a function of the peak density $n_0$. We measure the two-level response (red diamonds), the EIT response (blue squares) and the one with pre-excited impurities (green circles). The absorption is averaged over an area of 3x3 pixels at the center of the cloud. Error bars represent the measured standard deviation obtained from 10 repetitions. The data is compared to a hard-sphere model developed in section~\ref{sec:ModelIEI}: the absorption is calculated from the two-level susceptibility $\chi_{2l}^{~}$ (red dotted line), the EIT susceptibility $\chi_{eit}^{~}$ (blue dashed line), and the susceptibility with impurities $\chi_{imp}^{~}$ (green solid line). 
{\bf(b)} Comparison between measured and predicted SNRs. Measurements are performed with an exposure time $t_{exp}=10\,\rm{\mu s}$. The predicted SNR is calculated assuming photon and camera noise (see sec.~\ref{sec:detection}) as well as atomic density fluctuations, measured to be on the level of $8\%$. The SNR of the measurements is slightly lower than predicted, probably because of additional noise sources.}
\label{fig:ModelDensityComparison}       
\end{figure}

We first measure the two-level response of the ground state atoms by turning off the coupling light field (red diamonds in fig.~\ref{fig:ModelDensityComparison}a). The measured absorption $A_{2l}$ allows us to extract the peak density of the cloud using eq.~\eqref{eq:chi_2lvl}, taking into account the small saturation effect due to the non-zero probe Rabi frequency $\Omega_p=2\pi \cdot (1.14\pm0.02)\,\rm{MHz}$, which leads to a correction of about $7\%$. 
Since it is used to calibrate the peak density axes, the data coincides exactly with the theoretical two-level absorption (red dotted line).

We then turn on the coupling light field to reach the conditions for EIT and perform the same acquisition (blue squares). We analyse only the probe absorption $A_{eit}$ in an area corresponding to the center of the coupling laser beam, which allows us to assume $\Omega_c$ as constant. Independent measurements relying on Autler-Townes spectra are used to calibrate the strength of the coupling $\Omega_c=2\pi \cdot (9.45\pm 0.74)\,\mega\hertz$. The observed loss of transparency when the atomic density increases is the signature of van der Waals interaction between probe atoms. The data is compared to the hard-sphere model (blue dashed line) using eq.~\eqref{eq:chi_eit}. We obtain good agreement assuming a Rydberg blockade of $R_{rr}=3.6\pm 0.3\,\micro\meter$, while the theoretical prediction is $2.5\,\micro\meter$, meaning that the effect of interactions appears to be stronger than predicted from the dipole moment of the state used. In these conditions the blockaded fraction $f_{rr}$ changes with the density from 0 to $0.31\pm 0.03$. 

In the last realization we pre-excite some impurities during a fixed excitation time $t_{exc}=5\,\micro\second$ and we record the optical response (green circles). The larger absorption compared to the previous case is caused by additional absorbing probe atoms due to the presence of impurities. The evolution of the optical response measured from the absorption $A_{imp}$ is again compared to the hard-sphere model using eq.~\eqref{eq:chi_IMP}. The anisotropic interaction strength is estimated to be $C_3^{ir}=2\pi \cdot 1.7\,\rm{GHz}\cdot\rm{\mu m^3}$ considering an angular averaging~\cite{walker2008}, leading to a blockade radius of $R_{ir} = 6.1\,\micro\meter$, roughly 2 times larger than $R_{rr}$. The model reproduces very well the data assuming a constant fraction of volume blockaded by the impurities of $f_{ip}=0.24 \pm 0.02$, resulting in $26 \pm 5$ Rydberg impurities embedded in the whole sample. From the field ionization we estimate the number of impurities to be $\sim 12$, which is consistent within the large uncertainty of the ion detection efficiency.  

The excellent agreement with the data justifies the use of a hard sphere model to describe IEI. In the density range we have explored, figure~\ref{fig:ModelDensityComparison}a shows that the signal increases with the atomic peak density, indicating that the amplification factor, expressed as the number of additional absorbers per impurity, is also raising. Nevertheless we expect that at higher densities the signal will decrease due to the incrementing impact of probe-probe interactions, as simulated in fig.~\ref{fig:LightPropNoiseSNR}c. 

Using the measured data and the results from the model it is possible to compare the measured and predicted SNRs defined in eq.~\eqref{eq:SNR} (fig.~\ref{fig:ModelDensityComparison}b). The atomic density fluctuations are estimated from the two-level response to be on the level of $8\%$ and do not contribute much to the total noise. The agreement is qualitatively good, but the measured SNR is smaller than expected and it suggests that additional noise sources present in the experiment must be considered.

\subsection{Imaging few P-state impurities}
\label{sec:IEI_fewP}

After testing the model with experiments in the reservoir dipole trap, we perform a new set of IEI experiments in the tight optical dipole trap (see section \ref{subsec:SetupTightTrap}). Due to its restricted volume, diffusion of the Rydberg impurities out of this volume is negligible~\cite{Guenter2013}. Moreover the maximum number of impurities remains relatively small due to the blockade effect, which allows us to investigate the detection up to $\sim 10$ impurities in the sample.

We prepare the atoms in the ground state $\ket{g}$ at a temperature of $T = 1\,\rm{\mu K}$ with a peak density of $n_0=(1.4\pm0.3)\cdot 10^{11}\,\centi\meter^{-3}$ . The average number of impurities $N_i$ is changed by varying the duration time $t_{exc}$ of the three-photon excitation and can be separately determined via field-ionization detection (sec. \ref{subsec:SetupExcDet}). For each $t_{exc}$ we record the optical response under EIT conditions and compare it to the one in absence of impurities to deduce the additional absorption $A_{add}$. In figure~\ref{fig:additional_absorption}a we plot the number of additional absorbers $N_{add}$ for each pixel of the CCD camera, corresponding to an area of $a_{px}=4.3\,\micro\meter^2$ in the object plane, given by
\begin{equation}
	\label{eq:additional_absorbers}
	N_{add}=\frac{a_{px}(1+s_0)}{\sigma_0}  \Delta \OD = \frac{a_{px}(1+s_0)}{\sigma_0} \ln\left[(1-A_{add})^{-1})\right],
\end{equation}
where $s_0 = 2\Omega_p^2/\Gamma_e^2$ is the saturation parameter of the probe transition. This equation converts the measured signal into an effective number $N_{add}$ of atoms that would be absorptive instead of fully transparent due to the presence of impurities. The integrated number of additional absorbers along the vertical axis is plotted in figure~\ref{fig:additional_absorption}b (red solid line) and is compared to a Gaussian fit (blue dashed line) where the scale of the different plots has been fixed. This representation helps to observe how large the signal is compared to the residual noise in the region outside of the cloud.

\begin{figure}[t]
\centering
\includegraphics[width=\linewidth,keepaspectratio=true]{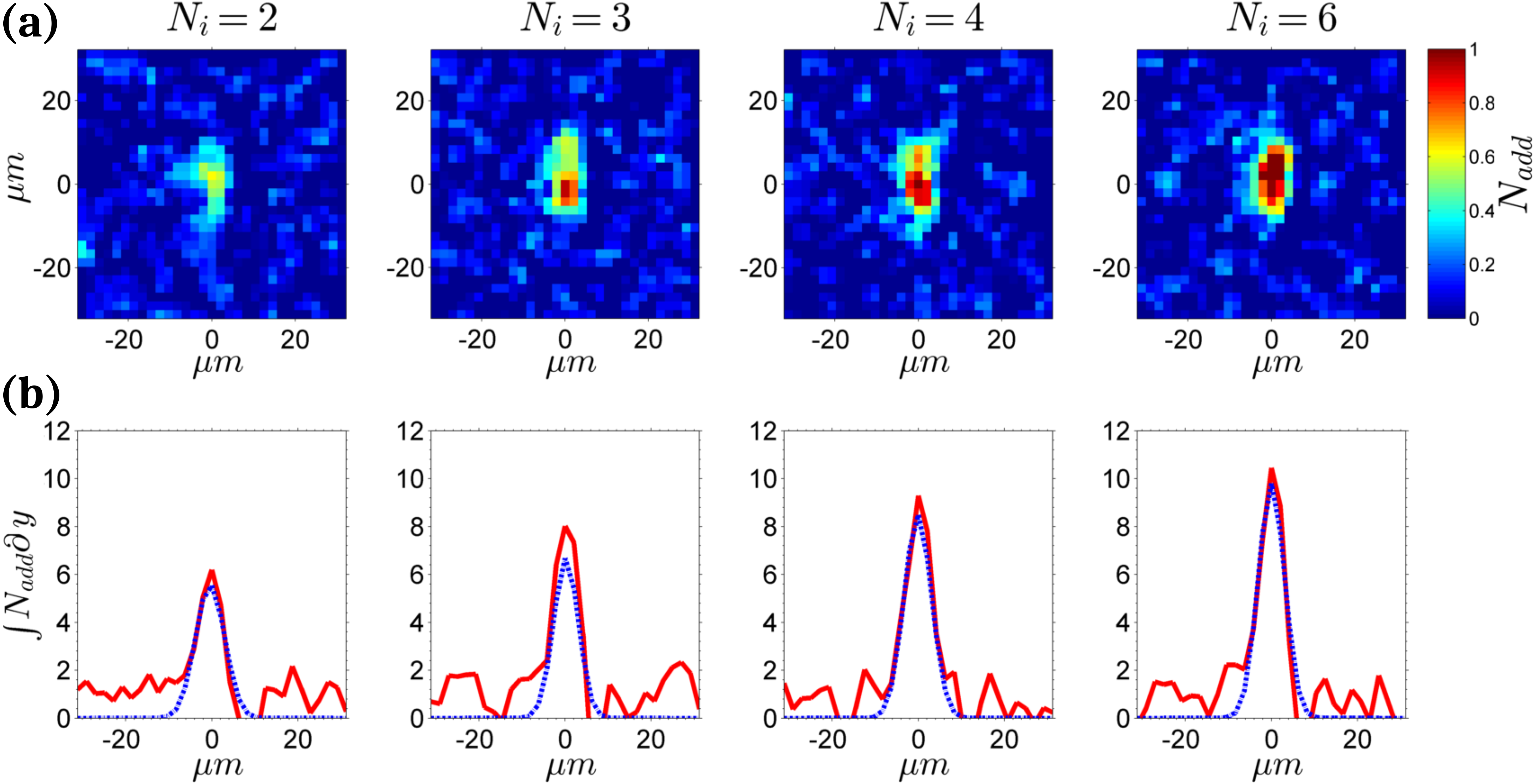}
\caption{Interaction-enhanced imaging of $42P_{3/2}$ Rydberg atoms in a small atomic sample. {\bf(a)} Spatial distribution of the additional absorbers per pixel $N_{add}$ for increasing average number of impurities $N_i=2,\,3,\,4,\,6$ as measured by field ionization. Each measurement is averaged over 50 repetitions with an exposure time of $t_{exp}=30\,\rm{\mu s}$. {\bf(b)} Integral along the picture's vertical direction
of the measured number of additional absorbers (solid red) and of a 2D Gaussian fit to the data (blue dashed line). The fit integral allows us to extract $\Sigma N_{add}$ from the data, while removing the noise outside of the sample region.}
\label{fig:additional_absorption}
\end{figure}

In order to distinguish the signal coming from different numbers of impurities, we have increased the EIT bandwidth $\sigma_{eit}$ by strongly increasing the coupling Rabi frequency $\Omega_c$. Because of the small size of the sample compared to the extension of the coupling beam, we assume $\Omega_c$ to be constant, equal to $26.2 \pm 2.3\,\mega\hertz$ according to independent measurements~\cite{Gavryusev2016}. This choice is relevant since we clearly observe in fig.~\ref{fig:additional_absorption} that the cloud becomes progressively more absorptive while increasing $N_i$. To characterize how the impurities affect the optical response, we extract the total number of additional absorbers $\Sigma N_{add}$ for each realization through a 2D Gaussian fit over the absorption area, which allows to remove the contribution of noise outside of the sample region. The plot in figure~\ref{fig:absorbers_impurities} shows the evolution of $\Sigma N_{add}$ with respect to the mean number of impurities $N_i$ measured by field ionization (blue points). It can be seen that $\Sigma N_{add}$ linearly increases for small $N_i$ before reaching a saturated value of $\sim 40$ additional absorbers for $N_i > 4$. 

Within the large Rabi coupling frequency $\Omega_c$ regime in which the experiment is performed, we can reasonably assume that $\chi_{eit}^0 \approx 0$ and we can also neglect the correction due to probe-probe interactions in equation~\eqref{eq:signal_simplified} since $f_{rr} \approx 0.02 \ll 1$. We can then simply express the total number of additional absorbers as $\Sigma N_{add} = f_{ir} N$ where $N$ is the total number of ground-state atoms in the sample. This formula shows that under our assumptions any atom within the impurity blockade volume will behave as an additional absorber. Since $\Sigma N_{add}$ saturates around $40$ atoms while $N$ is on the order of 400 in the experiment, $f_{ir}$ remains smaller than $10\%$, meaning that only part of the total volume is blockaded by the impurity-probe interactions. We can then express $\Sigma N_{add}$ in an even simpler form $\Sigma N_{add} = N_i \cdot n_0 V_{bl,i}$ (see sec.~\ref{subsec:modelHS}).

Hence our model predicts a linear dependence of the total number of additional absorbers with the number of impurities. The IEI amplification factor, as defined in sec.~\ref{sec:intro}, would be given here by $n_0 V_{bl,i}$. This prediction agrees well with the experimental data as shown in fig.~\ref{fig:absorbers_impurities}. The observed deviation at larger number of impurities might be interpreted as a Rydberg blockade effect between the impurities during their excitation, that we estimate to be significant above $\sim 10$ impurities. Impurities cannot be excited in the center of the sample, while it would be still possible in the tails, where the density is low and thus the change in the optical response negligible. Since in our case $R_{ii} \gg R_{ir}$, only a small fraction of the total volume is effectively blockaded.

\begin{figure}[t]
\centering
\includegraphics[width=0.75\linewidth,keepaspectratio=true]{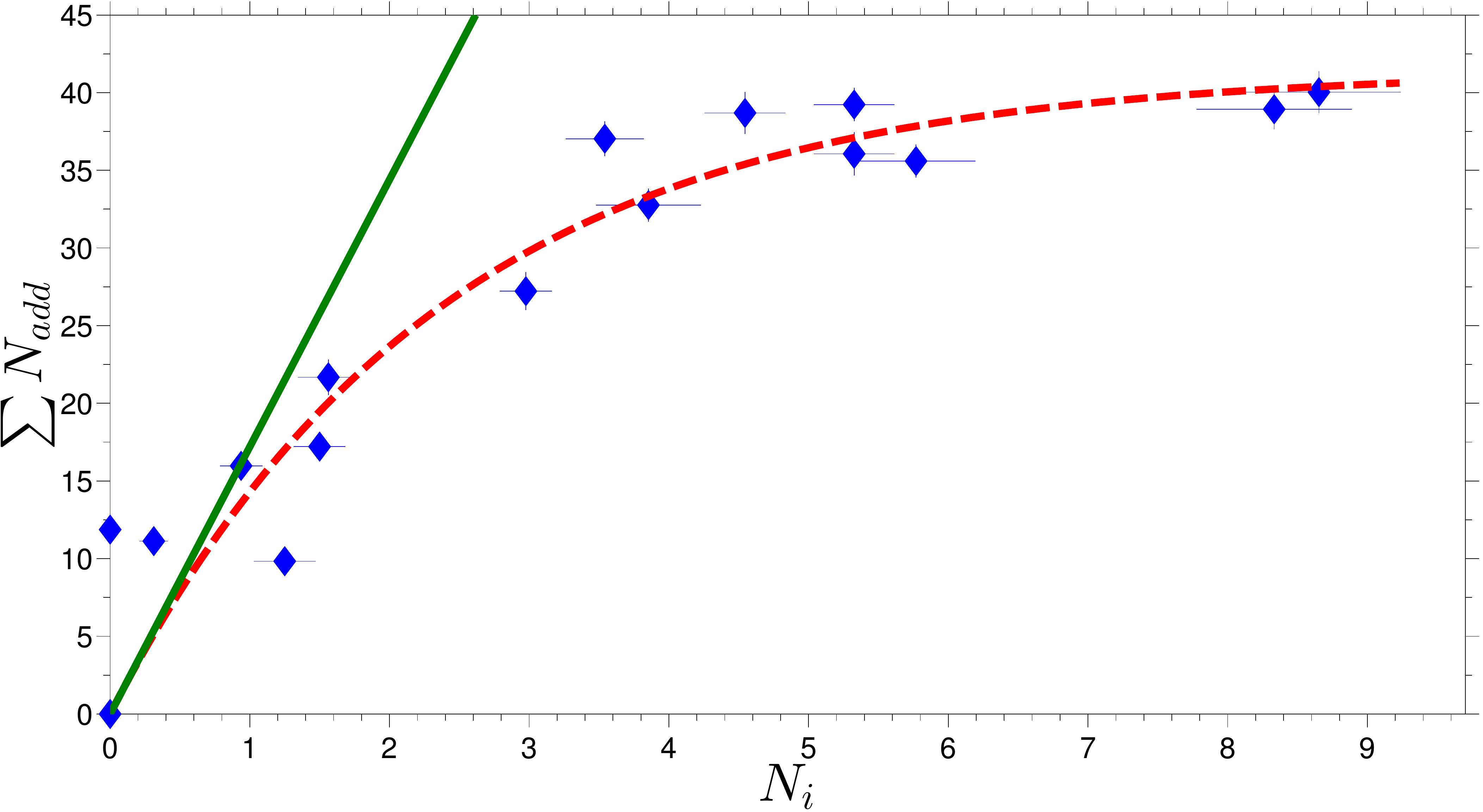}
\caption{Number of total additional absorbers $\Sigma N_{add}$ as a function of the mean number of impurities $N_i$. We first observe a nearly linear growth of the additional absorbers as expected from the hard-sphere model. Then the growth exponentially decays, reaching a saturated level of around 40 absorbers, that we interpret as a signature of the Rydberg blockade effect between the impurities. We fit the experimental data with an exponential rate model, that gives an initial slope of $17.4 \pm 0.2$, in excellent agreement with our model that predicts an amplification factor of 17.1. Nevertheless the precision of our measurement is lower than this agreement since the detection efficiency of the field-ionization is known with a large uncertainty. The errors are the standard errors of the mean.}
\label{fig:absorbers_impurities}
\end{figure}

These measurements demonstrate that IEI is suitable to detect individual Rydberg impurities with a sensitivity close to 1 impurity on average. Moreover in the regime of strong EIT coupling and few impurities, the number of additional absorbers changes linearly with the number of impurities and depends only on the atomic density $n_0$ and the EIT bandwidth $\sigma_{eit}$ through the blockade volume. This relationship allows to directly estimate the number of impurities without considering the microscopic details of the imaging.

\section{Outlook: conclusions and prospects}
\label{sec:outlook}

In this work we have performed optical imaging of Rydberg $P$ states using a detection technique called Interaction Enhanced Imaging, which consists in mapping the presence of Rydberg atoms on the optical response of an atomic medium. In order to prepare the atom sample, we have implemented and characterized a well-controlled excitation scheme that allows us to reliably prepare a chosen average number of atoms in a Rydberg $P$ state, without perturbing the cloud. The scheme is based on a three-photon off-resonant excitation which is convenient for preparation of $P$ states independently of any other Rydberg state. A theoretical analysis has allowed us to introduce a simple description of the excitation scheme in terms of an effective two-level system which provides accessible tools to fully optimize the excitation dynamics.

To gain insight into the imaging process, we have extended an analytic hard-sphere model, first introduced to describe interactions within the EIT medium, to include Rydberg impurity-medium interactions and we have used it to predict the optical response of the system and estimate the SNR. Then, using our technique, we have successfully imaged Rydberg atoms prepared in the state $\ket{42P}$ in a large atomic sample and we have compared the estimated signal with experimental data, finding good agreement. Furthermore measurements in a small atomic sample with dimensions comparable to the impurity-impurity blockade radius $R_{ii}$ have shown that we can reach a sensitivity of few Rydberg atoms with IEI, close to the best competing methods~\cite{Schwarzkopf2011,Schauss2012,McQuillen2013,Lochead2013}, with the advantage that we do not destroy the Rydberg atoms under observation.

A further improvement of IEI will consist in reaching single-impurity sensitivity on single-shot measurements. It would allow new studies of energy transport dynamics induced by dipolar interactions, e.g by following the dynamics of a single Rydberg impurity in a system with highly controllable geometry and dimensionality. Such sensitivity requires to enhance the SNR far above 1, while probing the system within a time short compared to the dynamics of the impurities, typically few $\mu$s. Leveraging the hard-sphere model predictions we aim to achieve a high SNR by magnifying the amplification factor by working with high probe atom densities and states with larger principle quantum numbers, as they exhibit stronger interaction and longer lifetimes. Additionally the imaging noise should be minimized to its lowest possible level. First analysis already shows strong noise reduction by suppressing atom density and light intensity fluctuations, but further work must be done to reach the photon shot noise limit. 

One of the main advantages of the IEI method is that the spatial resolution is given by the impurity-probe blockade radius $R_{ir}$ convolved with the optical resolution of the imaging system. Consequently it can be engineered to be smaller than the typical impurity-impurity distance $R_{ii}$, allowing for spatially-resolved single-impurity imaging. Combined with a high sensitivity the IEI technique might be promising for spatial correlation measurements of Rydberg ensembles in bulk atomic gases, opening perspectives to study energy transport mechanisms in open systems~\cite{Schonleber2015,Schempp2015} or the formation of Rydberg aggregates~\cite{Malossi2014,Schempp2014,Urvoy2015}.

\section*{Acknowledgements}
We thank G. G\"{u}nter and S. Whitlock for important contributions to the experiment planning and to the theoretical description of EIT in Rydberg atomic gases. We also thank R. F. Alves for help with numerical simulations, J. Evers and G. Faraoni for fruitful discussions and M. Weidem\"{u}ller for guidance. This work is supported in part by the Heidelberg Center for Quantum Dynamics, the Deutsche Forschungsgemeinschaft under WE2661/10.2 and the European Union H2020 FET Proactive project RySQ (grant N. 640378). M. Ferreira-Cao and V. Gavryusev acknowledge support from the EU Marie-Curie program ITN COHERENCE (Grant No. FP7-PEOPLE-2010-ITN-265031) and from the IMPRS-QD.

\bibliographystyle{epj}
\bibliography{paperEpjSt}

\end{document}